\documentclass{article}
\usepackage{graphicx} 

\usepackage{authblk}
\usepackage[english]{babel}

\usepackage[bottom]{footmisc} 

\usepackage[letterpaper,top=2cm,bottom=2cm,left=3cm,right=3cm,marginparwidth=1.75cm]{geometry}

\usepackage{booktabs}
\usepackage{siunitx}  
\usepackage{amsmath}
\usepackage{amsthm}
\usepackage{amssymb}
\usepackage{graphicx}
\usepackage{enumitem}
\usepackage{mathrsfs}
\usepackage{tikz}
\usetikzlibrary{shapes.geometric} 

\usepackage{subcaption}
\usepackage{lipsum}
\usepackage{dsfont}
\usepackage{multirow}
\usepackage{xcolor}

\usepackage{xurl}

\usepackage[breaklinks=true]{hyperref}

\usepackage[ruled,vlined]{algorithm2e}
\usepackage{tikz}
\usetikzlibrary{shapes.geometric, arrows, positioning}

\newcommand{\reddiamond}{%
  \tikz[baseline=-0.5ex] \node[
    shape=diamond,        
    fill=red,             
    draw=none,            
    minimum size=8pt,     
    inner sep=0pt         
  ] {};%
}

\def \k {\kappa}
\def \K {\mathcal{K}}

\def \x {\mathbf{x}}
\def \y {\mathbf{y}}
\def \z {\mathbf{z}}
\def \C {\mathcal{C}}
\def \D {\mathcal{D}}
\def \E {\mathbb{E}}
\def \F {\mathcal{F}}
\def \H {\mathcal{H}}
\def \P {\mathcal{P}}

\def \R {\mathbb{R}}

\newtheorem{definition}{Definition}

\newcommand{\tabincell}[2]{\begin{tabular}{@{}#1@{}}#2\end{tabular}}

\DeclareMathOperator*{\argmin}{argmin}
\DeclareMathOperator*{\argmax}{argmax}

\title{A New Framework for Explainable Rare Cell Identification in Single-Cell Transcriptomics Data}

\usepackage{hyperref}
\hypersetup{
    colorlinks=true,       
    linkcolor=blue,        
    citecolor=blue,       
    filecolor=blue,     
    urlcolor=blue,         
    pdftitle={A New Framework for Explainable Rare Cell Identification},
    pdfauthor={Di Su, et al.},
    bookmarksopen=true,    
}

\usepackage{authblk}

\author{Di Su}
\author{Kai Ming Ting\thanks{Corresponding author: \href{mailto:tingkm@nju.edu.cn}{tingkm@nju.edu.cn}}}
\author{Jie Zhang}
\author{Xiaorui Zhang}
\author{Xinpeng Li}

\affil{State Key Laboratory for Novel Software Technology, Nanjing University}

\date{}

\begin{document}
\maketitle

\begin{abstract}
The detection of rare cell types in single-cell transcriptomics data is crucial for elucidating disease pathogenesis and tissue development dynamics.
However, a critical gap that persists in current methods is their inability to provide an explanation based on genes for each cell they have detected as rare.
We identify three primary sources of this deficiency. 
First, the anomaly detectors often function as ``black boxes'', designed to detect anomalies but unable to explain why a cell is anomalous.
Second, the standard analytical framework hinders interpretability by relying on dimensionality reduction techniques, such as Principal Component Analysis (PCA), which transform meaningful gene expression data into abstract, uninterpretable features.
Finally, existing explanation algorithms cannot be readily applied to this domain, as single-cell data is characterized by high dimensionality, noise, and substantial sparsity.

To overcome these limitations, we introduce a framework for explainable anomaly detection in single-cell transcriptomics data which not only identifies individual anomalies, but also provides a visual explanation based on genes that makes an instance anomalous.
    This framework has two key ingredients that are not existed in current methods applied in this domain. First, it eliminates the PCA step which is deemed to be an essential component in previous studies. 
    Second, it employs the state-of-art anomaly detector and explainer as the efficient and effective means to find each rare cell and the relevant  gene subspace in order to provide explanations for each rare cell as well as the typical normal cell associated with the rare cell's closest normal cells.
\end{abstract}

\section{Background}

Anomaly detection has long been one of the key research areas in machine learning \cite{chandola2009anomaly,pang2021deep}.
In general terms, the goal is to find patterns in the data that deviate from the expected normal behavior \cite{Foorthuis2021}. This capability is critical in diverse fields.
In parallel, the rise of explainable artificial intelligence (XAI) has highlighted the need to move beyond ``black-box" predictions, demanding that models provide explanations for their outputs \cite{arrieta2020explainable,Rai2020,Hassija2024}.
The integration of these two fields, extending from detecting an anomaly to explaining its anomalous nature, is crucial in real-world applications. 

In the domain of single-cell transcriptomics, the identification of rare cell types is an important step in understanding disease mechanisms and the cellular processes that govern tissue development. 
This challenge has prompted the development of various computational methods aimed at automating this process~\cite{zhang2019scina, shao2020sccatch,deng2019scalable}.
Indeed, developing advanced computational methods to decipher biological structures from such data is a vibrant research area, spanning from the scalable discovery of complex spatial domains in spatial transcriptomics \cite{zhang2025kernel} to the identification of rare cells in single-cell analysis.
While this task is naturally an anomaly detection problem \cite{Grün2015Single-cell},
a key issue in the process is the lack of a meaningful, gene-level explanation for detected anomalies, which is essential for elucidating the underlying mechanisms of cell-type-specific phenomena~\cite{brown2013integrative}.
This deficiency stems from at least three sources inherent in the field.

First, anomaly detection algorithms, ranging from classical methods to state-of-the-art approaches like scCAD \cite{Xu2024scCAD},
are primarily engineered for detection rather than explanation.
Their main function is to assign an anomaly score to each cell, leaving them incapable of elucidating the biological drivers of a cell's anomalous state.

Second, standard single-cell analysis pipelines, which typically involve dimensionality reduction, inherently hinder explainability \cite{luecken2019current, duo2020systematic}.
While these steps are widely considered an essential step for managing high-dimensional data \cite{He2022,xiang2021,Wolf2018}, to the extent that developing novel, specialized methods remains an active research frontier \cite{chen2025smopca},
they often create the primary hurdle to interpretability. 
Techniques like Principal Component Analysis (PCA) transform biologically meaningful gene expression data into abstract features. 
These features, being linear combinations of original genes, lack direct biological interpretability.

Third, existing algorithms designed for anomaly explanation are unable to deal with the unique characteristics of single-cell transcriptomics.
These methods, typically developed for clean and low-dimensional data, are hampered by the high dimensionality, noise, and substantial sparsity inherent in single-cell data \cite{Chen2019,Lahnemann2020}. 
For instance, the pipeline proposed by Samariya et al. \cite{samariya2023explanation}, while effective at identifying explanatory subspaces for anomalies in healthcare data, is not directly transferable to the challenging landscape of single-cell transcriptomics.

In this work, we present a unified framework that overcomes these limitations by specifically addressing the challenges of high dimensionality, data sparsity, and the need for intuitive anomaly explanation.

\section{Results}

Our main contributions are:

\begin{enumerate}
    \item \textbf{Establishing a PCA-Free Framework for High-Fidelity Rare Cell Detection:}
    We challenge the standard single-cell analysis pipeline by demonstrating that dimensionality reduction techniques, specifically PCA, act as an information bottleneck that hampers the detection of rare biological signals.
    \begin{itemize}
        \item We introduce a framework that operates directly on Highly Variable Genes (HVGs). By eliminating the PCA step (which is previously deemed essential in major studies \cite{He2022,xiang2021,Wolf2018}), we preserve the raw biological fidelity of the data.
        \item We employ the Isolation Distributional Kernel (IDK) \cite{Xu2020IDK} adapted for high-dimensional sparse data. We demonstrate that this direct high-dimensional detection strategy is superior to PCA-based methods in capturing subtle rare transcriptomic states that are often lost in compressed latent spaces.
    \end{itemize}

    \item \textbf{Pioneering ``Contrastive Gene-Subspace'' Explainability:}
     We adapted the SiNNe algorithm to generate instance-specific biological explanations, transforming a bare ``anomaly score'' into a biological narrative. We conduct extensive experiments to demonstrate that meaningful explanations can be generated using a subspace of genes that highlights:
    \begin{itemize}
        \item [(i)] the rarity of an anomaly with respect to all cells of a normal cluster which is closest to the anomaly;
        \item [(ii)] the characteristic genes of a typical normal cell with respect to the cluster of normal cells.
    \end{itemize}

    The experiments include a comparative analysis of different anomaly detectors, which show that our framework plays an essential role in determining their relative importance.

    \item \textbf{A Rigorous Protocol for Validating ``Group Anomalies'':}
    We demonstrate the framework's capability to validate and dissect results from other detectors. By applying our explanatory tool to the ``clustered anomalies'' identified by scCAD \cite{Xu2024scCAD}, we provide a method to distinguish genuine, biologically distinct rare populations from computational artifacts or homogenous clusters. This represents a novel application of XAI for verification in bioinformatics.
\end{enumerate}

Our work fills a critical gap in single-cell transcriptomics: the disconnect between detecting a rare cell and understanding its biological nature. While current methods rely on global cluster markers or uninterpretable embeddings, our framework provides a practical, gene-level solution. It enables researchers to answer why a specific cell is anomalous and how it deviates from its microenvironment, offering verifiability to anomaly detection tasks that was previously unattainable.





\subsection{Method Overview}

\begin{figure}[ht]
    \centering
    \includegraphics[width=\linewidth]{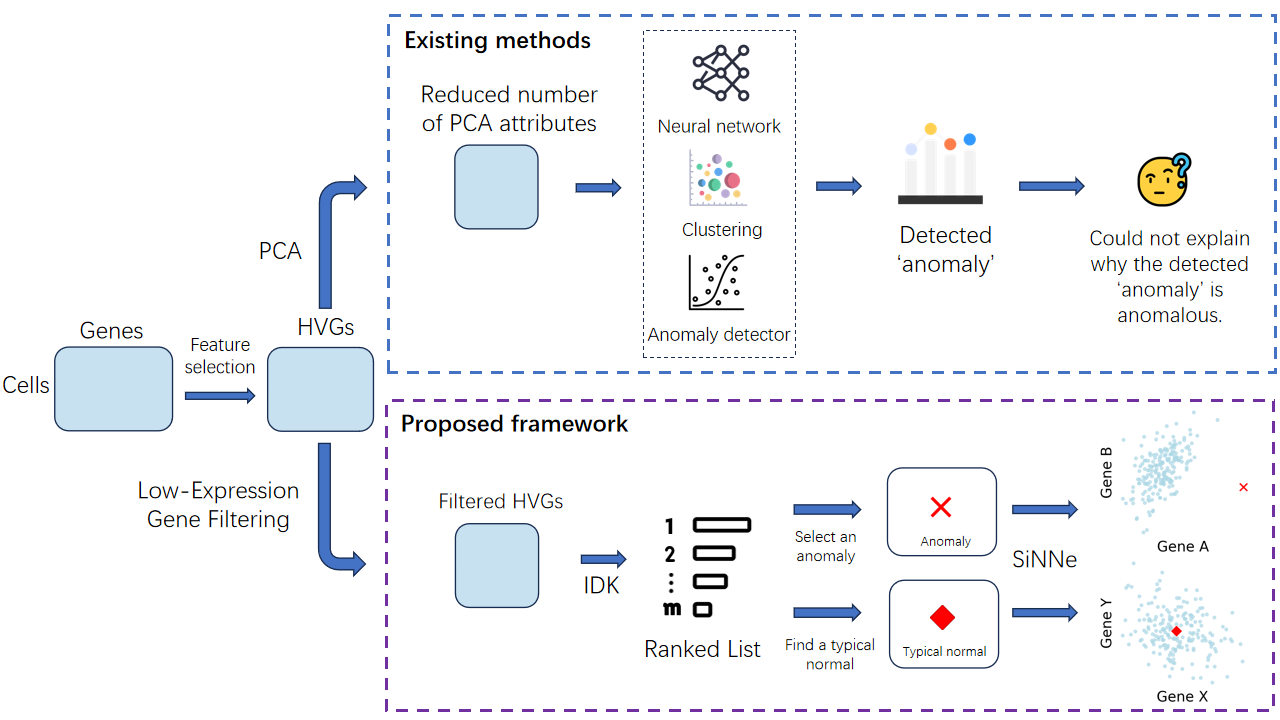}
    \caption{Overview of existing methods \cite{Grün2015Single-cell,Zhou2025AnomalGRN,Ferré2021Anomaly} and our proposed framework. In the proposed framework, we first select highly variable genes and compute anomaly scores using the Isolation Distributional Kernel (IDK) \cite{Xu2020IDK}, which measures each cell’s similarity to the dataset distribution.  We then rank cells and identify the top anomaly and typical normal. 
    Finally, we apply SiNNe \cite{samariya2020new} to extract concise gene subspaces explaining why a specific cell is anomalous or typically normal.}
    \label{fig:pipeline}
\end{figure}

Our proposed framework, illustrated in \autoref{fig:pipeline},
preprocesses the given  single-cell transcriptomics data by selecting highly variable genes (HVGs) only. Then, the HVG data is treated in two components: rare cell detection and rare cell explanation. The first component employs the Isolation Distributional Kernel (IDK) \cite{Xu2020IDK} to score each cell or spot.
IDK is a data-dependent kernel that is robust to high-dimensional and sparse data.
It effectively measures how anomalous a cell is relative to the distribution of all cells in the HVGs dataset.
For the subsequent explanatory component, SiNNe \cite{samariya2020new} is an algorithm that operates efficiently in high-dimensional spaces, but it is unable to deal with the original single-cell transcriptomics data because of many missing values. Our preprocessing has enabled HVGs data to be treated successfully by SiNNe. 
This is crucial, as it enables SiNNe to search for and identify the optimal gene subspace that distinguishes a target cell from a given set of neighbor cells.
Our framework generates explanations not only for each rare cell but also for the typical cell of a normal cluster which has cells that are closest to the rare cell. 

The original SiNNe was designed only to identify the gene subspace that makes a cell anomalous.
In our framework, our version can also find the subspace of genes that best characterizes why a cell is a typical.
This dual capability allows us to provide a contrastive explanation, highlighting both the genes driving a cell's rarity and the genes that explain why a cell is considered normal.

To ensure these gene-based explanations are easily interpretable, SiNNe is configured to find an optimal subspace with a maximum of three dimensions. 
The gene subspaces produced for both rare and typical normal cells can then be directly visualized, offering an intuitive graphical representation that shows their unique genetic signatures.



In the rest of the paper, the terms ``feature" and ``gene'' are synonymous, and they are used interchangeably; so as ``instance" and ``cell''. 
Furthermore, we treat the task of identifying rare cells as an anomaly detection problem.
Consequently, a ``rare cell" is considered an ``anomaly".

\subsection{Aims of the Experiments}
\label{sec-aims}
The following subsections aim to answer these questions:
\begin{itemize}[label={}, leftmargin=0pt, itemsep=0pt]
\item \textbf{(I) Which are the effective features for anomaly detection: HVG versus PCA?}
This aims to evaluate the relative effectiveness of using highly-variable genes (HVG) and PCA-derived features for anomaly detection.
\item \textbf{(II) How to explain an anomaly of being different from its most similar normal cells?}
This aims to find a way to explain each detected anomaly by identifying the genes which differentiate the detected anomaly from its closest normal cells.
\item \textbf{(III) How to characterize the normality of a typical cell?}
Given a set of normal cells, this explores a way to explain the characteristics of a typical normal cell with respect to the set.
\item \textbf{(IV) Do different anomaly detectors identify the same or different anomalies?}
An assessment method is employed to compare different detectors in a dataset to determine whether a single detector is sufficient or multiple detectors are required because they capture different types of anomalies.
\item \textbf{(V) Are the ``anomaly clusters” detected by scCAD \cite{Xu2024scCAD} explainable biological entities?}
This question investigates the biological entities of the anomaly clusters detected by scCAD and explores how these clusters manifest as anomalies within their respective low-dimensional subspaces.
\end{itemize}
We will answer the first question, the next two questions, the fourth question, and the last question in the following four subsections.

\subsection{Direct High-Dimensional Analysis Preserves Rare Signals Lost by Dimensionality Reduction}
\label{section:gamma}

Standard single-cell analysis pipelines rely heavily on dimensionality reduction techniques, such as Principal Component Analysis (PCA), to mitigate noise and computational complexity. However, PCA prioritizes axes of maximal global variance, which are typically dominated by abundant cell types. We hypothesize that this process acts as an information bottleneck: the subtle, high-dimensional gene expression signatures that characterize rare cells are often compressed or discarded because they contribute little to the global variance.

To validate this hypothesis and determine whether the proposed direct high-dimensional analysis (using Highly Variable Genes, HVG) captures biological signals lost by PCA, we compared the two representations using the $\Gamma$ metric \cite{Ting2025}.

Unlike simple overlap measures, the $\Gamma$ metric assesses the comprehensiveness of a feature set. Specifically, $\Gamma(\text{PCA}|\text{HVG})$ measures the proportion of the population that must be examined using PCA features to recover the anomalies identified by HVG features. A high value implies that anomalies clear to HVG are buried deep in the PCA ranking.


Our analysis, conducted with the Isolation Distributional Kernel (IDK) detector, revealed a clear asymmetry across all four datasets (\autoref{tab:feature_comparison}). We consistently observed that \( \Gamma(\text{PCA}\!\mid\!\text{HVG}) \gg \Gamma(\text{HVG}\!\mid\!\text{PCA})\). This result indicates that HVG features are far more comprehensive for finding rare cells than PCA features.

\begin{table}[ht]
  \centering 
  \caption{Comparing the relative goodness of the feature sets derived from PCA and HVG using the \(\Gamma(\cdot\!\mid\!\cdot)\) metric on the top \(\varepsilon=0.05\) fraction of anomalies.}
  \label{tab:feature_comparison}
  \sisetup{table-format=1.2} 
  \begin{tabular}{l|S|S|l}
    \toprule
    \textbf{Dataset} & {\textbf{$\Gamma$(PCA$|$HVG)}} & {\textbf{$\Gamma$(HVG$|$PCA)}} & \textbf{Comment}\\
    \midrule
    Tutorial  & 0.82 & 0.33 & \(\Gamma(PCA\!\mid\!HVG) \gg \Gamma(HVG\!\mid\!PCA) \approx 0.33\) \\
    Airway          & 0.94 & 0.50 & \(\Gamma(PCA\!\mid\!HVG) \gg \Gamma(HVG\!\mid\!PCA) \approx 0.5\) \\
    Tonsil          & 0.77 & 0.67 & \(\Gamma(PCA\!\mid\!HVG) \gg \Gamma(HVG\!\mid\!PCA) \gg 0.5\) \\
    Crohn           & 0.51 & 0.19  & \(\Gamma(PCA\!\mid\!HVG) \gg \Gamma(HVG\!\mid\!PCA) \approx 0.19\)\\
    \bottomrule
  \end{tabular}
\end{table}


The only notable observation is on the Tonsil dataset, where both metrics achieve high scores.
Nevertheless, the asymmetric relationship \(\Gamma(PCA\!\mid\!HVG) \gg \Gamma(HVG\!\mid\!PCA)\) still holds, confirming the superiority of HVG features.
Therefore, the HVG feature set was used for all subsequent experiments.

\subsection{Explaining an Anomaly and Characterizing a Normal Cell}
\label{sec:explain_A and Charactering_N}
To answer the second and third questions, we conduct experiments on several datasets.
We first use the Tutorial dataset to demonstrate how to explain individual anomalies and characterize typical normal cells.
We then showcase how the framework adapts to more complex scenarios.

\subsubsection{Tutorial Dataset }
\label{sec-tutorial}

\begin{figure}[t]
  \centering
  \begin{subfigure}[t]{0.24\textwidth}
    \includegraphics[width=\textwidth]{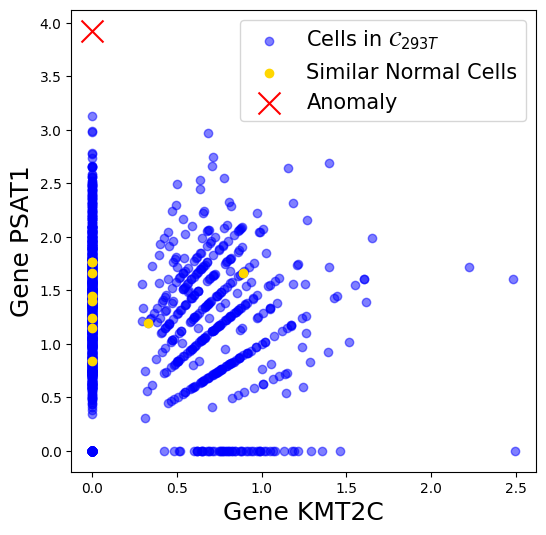}
    \caption{$\textbf{a}^{(1)}$ in $\mathcal{C}_{293T}$}
    \label{fig:tutorial_q68}
  \end{subfigure}
  \hfill
  \begin{subfigure}[t]{0.24\textwidth}
    \includegraphics[width=\textwidth]{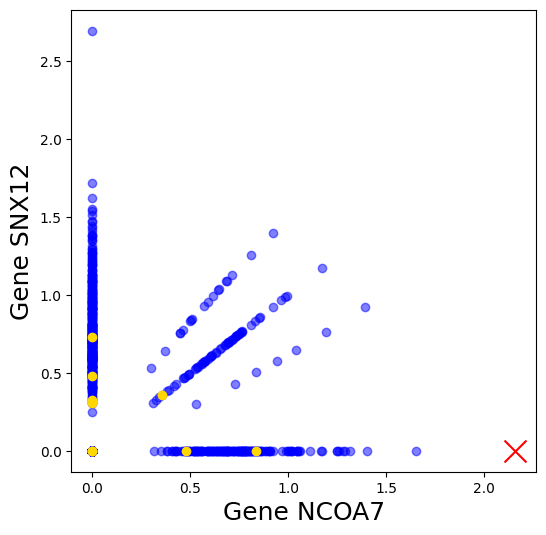}
    \caption{$\textbf{a}^{(2)}$ in $\mathcal{C}_{293T}$}
    \label{fig:tutorial_q1504}
  \end{subfigure}
  \hfill
  \begin{subfigure}[t]{0.24\textwidth}
    \includegraphics[width=\textwidth]{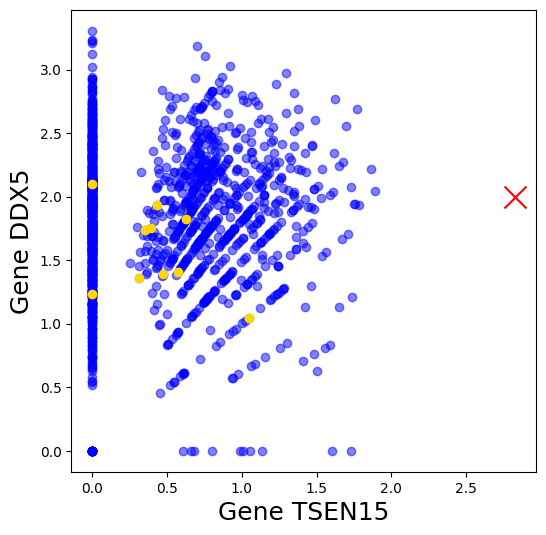}
    \caption{$\textbf{a}^{(3)}$ in $\mathcal{C}_{293T}$}
    \label{fig:tutorial_q1510}
  \end{subfigure}
  \hfill
  \begin{subfigure}[t]{0.24\textwidth}
    \includegraphics[width=\textwidth]{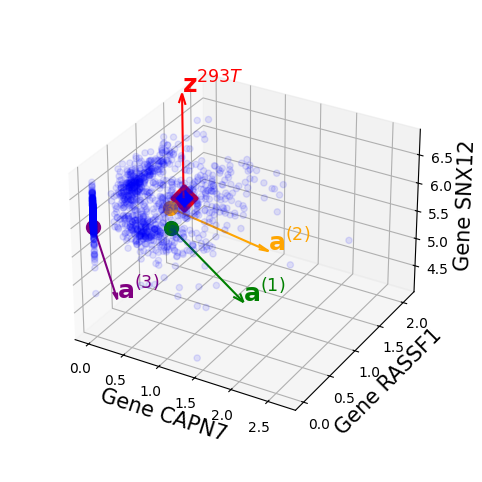}
    \caption{$\textbf{z}^{{293T}}$}
    \label{fig:tutorial_q_normal}
  \end{subfigure}
  \caption{Visualization of SiNNe‐derived feature subspaces in the Tutorial dataset.
  Anomalous cells are shown as ``\textcolor{red}{x}'' in subfigures (a-c);
  and the typical normal instance $\textbf{z}^{293T}$ is shown as ``\protect\reddiamond" in subfigure (d), which is compared against all cells in $\mathcal{C}_{293T}$ (blue dots) and the three detected anomalies.
  }
  \label{fig:tutorial_explanations}
\end{figure}

IDK identifies the following instances in the dataset: I\#68, I\#1504 and I\#1510 as the top three anomalies; and their top 10 normal neighbors are all members of the same cell type labeled 293T.
This scenario exemplifies what we define as an intra-type anomaly: an anomaly which is only relative to its own annotated cell type.
Let the set of cells with cell type 293T be
\[
\mathcal{C}_{293T} = \{\,x \in D \mid \text{Cell-Type}(x)=\text{293T}\}.
\]

\paragraph{Explaining Anomalies}
For each of the three anomalies, SiNNe identifies the genes that differentiate it from the rest of the instances in $\mathcal{C}_{293T}$. 
\autoref{fig:tutorial_q68}, \ref{fig:tutorial_q1504} and \ref{fig:tutorial_q1510} present the SiNNe‑selected gene subspaces for the anomalous cells.
We have the following observations:

\begin{itemize}
  \item \textbf{$\textbf{a}^{(1)}$ I\#68}: \autoref{fig:tutorial_q68} shows that genes \emph{KMT2C} and \emph{PSAT1} distinguish $\textbf{a}^{(1)}$ from all other cells in $\mathcal{C}_{293T}$. 
  \item \textbf{$\textbf{a}^{(2)}$ I\#1504}: \autoref{fig:tutorial_q1504} shows that genes \emph{NCOA7} and \emph{SNX12} differentiate $\textbf{a}^{(2)}$ from all the other cells of $\mathcal{C}_{293T}$.  
  \item \textbf{$\textbf{a}^{(3)}$ I\#1510}: \autoref{fig:tutorial_q1510} shows that genes \emph{TSEN15} and \emph{DDX5} explain why $\textbf{a}^{(3)}$ is anomalous.  
\end{itemize}
Distinct genes are used to explain each of the three anomalies, revealing that their anomalous status is characterized by different biological characteristics.

\paragraph{Characterizing a Typical Normal Cell}
Beyond explaining anomalies, our framework can also characterize the nature of a typical normal cell, directly answering the third question of our study.
To do this, we first identify the most representative instance from the normal cells that are labeled `293T', \textit{i.e.}, $\textbf{z}^{293T}$ (I\#23), which is the cell with the highest similarity to all other cells within $\mathcal{C}_{293T}$.

We then use SiNNe in its normal explanation mode to find the gene subspace that best describes what makes $\textbf{z}^{293T}$ typical. 

\autoref{fig:tutorial_q_normal} shows the distribution of all members of the cell type labeled 293T using the three genes identified by SiNNe to characterize the typical normal cell of this cluster.
The three genes are CAPN7, RASSF1 and SNX12.
The typical normal cell is I\#23 ($\textbf{z}^{293T}$).
All members of $\mathcal{C}_{293T}$, including the three anomalies, have similar characteristics based on these three genes.

It should be noted that the dataset comprises 1540 cells labeled 293T, and 16 cells labeled Jurkat.  
The 16 Jurkat cells obtain global anomaly ranks between 74 and 141 out of 1556 instances, which could not be considered as top anomalies, as none of the top 50 anomalies detected by IDK belongs to Jurkat.

\paragraph{Do Anomalies with Similar Scores Have Similar Characteristics?}

In the Tutorial dataset, we observed a notable phenomenon:
the IDK detector assigned identical, minimal similarity scores (zero) to approximately 100 cells.
This raises a question: do these top-ranking anomalies, which are indistinguishable by their anomaly scores, share the same underlying biological characteristics?
Or do they represent distinct types of anomalies? To investigate this, we use the t-SNE \cite{van2008visualizing} visualization together with our explanation method.

\begin{figure}[t]
  \centering
  \begin{subfigure}[b]{0.45\textwidth}
    \includegraphics[width=\textwidth]{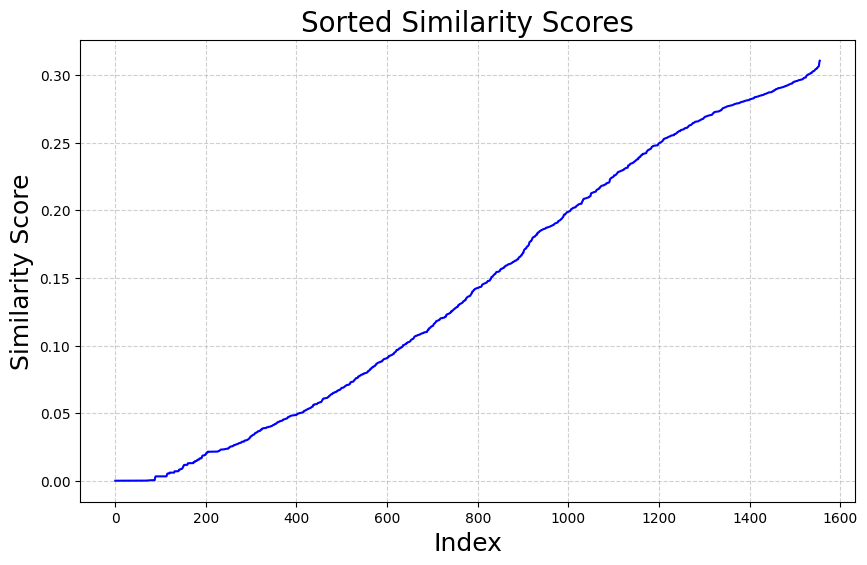}
    \caption{Plot of the sorted IDK similarity scores.}
    \label{fig:score_dist}
  \end{subfigure}
  \hfill
  \begin{subfigure}[b]{0.45\textwidth}
    \includegraphics[width=\textwidth]{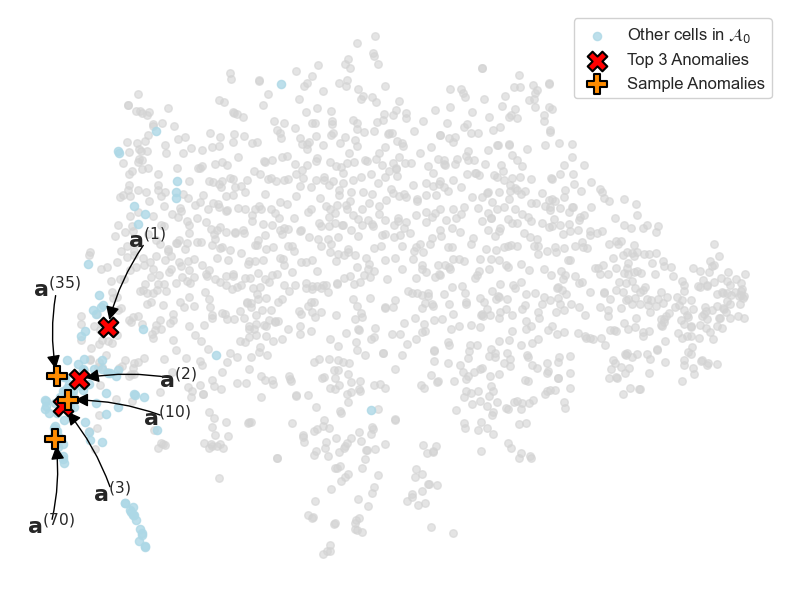}
    \caption{t-SNE projection of $\mathcal{A}_0$.}
    \label{fig:tsne_zero}
  \end{subfigure}
  \caption{Global similarity score distribution and spatial mapping of all instances on the t-SNE plot, with zero-score instances highlighted.}
\end{figure}

\textbf{Observation: A Cluster of Zero-Score Anomalies}
\autoref{fig:score_dist} presents the distribution of IDK similarity scores across all cells.
A significant number of instances ($\approx$ 100) have a score of exactly zero, indicating maximal dissimilarity from the dataset's overall distribution.
We denote the set of cells as $\mathcal{A}_0$.
To investigate the relationship between $\mathcal{A}_0$ and the rest of the population, we projected the entire dataset into a 2D space using t-SNE.
As shown in \autoref{fig:tsne_zero}, the members of $\mathcal{A}_0$ are not randomly scattered but instead exhibit a significant degree of co-localization, suggesting that they might form a single, coherent cluster of anomalies.

\textbf{Hypothesis and Experimental Design}
The proximity of these zero-score anomalies in the t-SNE plot suggests that they are in a homogeneous group.
However, we hypothesize that this apparent clustering is a misleading artifact of the dimensionality reduction process  employed in t-SNE, rather than a reflection of true biological similarity.

Our intuition is grounded in the known behavior of t-SNE when dealing with sparse, high-dimensional data.
t-SNE's primary objective is to preserve the local structure of dense neighborhoods.
Anomalies are rare and isolated instances that lack such neighborhoods.
Therefore, these instances are a collection of cells identified as anomalous for different biological reasons, and their true nature can only be revealed through individual explanations, rather than a global visualization.

Our previous analysis revealed that the top-3 anomalies are distinct from the others. To further test the heterogeneity of the entire zero-score population, we conducted another experiment.
Instead of re-examining the top-3 anomalies, we randomly sampled three additional anomalies within the zero-score set:
\(\textbf{a}^{(10)}\), \(\textbf{a}^{(35)}\) and \(\textbf{a}^{(70)}\).
Their locations based on the t-SNE projection are shown in \autoref{fig:tsne_zero}.
The experiment is to perform the same SiNNe analysis on these new anomalies as was done for the top-3:
we seek to find the gene subspace that distinguishes each of these anomalies individually from $\mathcal{C}_{293T}$.
If each requires a distinct set of genes for its explanation, it would provide evidence for the heterogeneity of the $\mathcal{A}_0$ population.

\textbf{Results: Distinct Explanations for Co-located Anomalies.}

\begin{figure}[ht]
\centering
\vspace{0.5cm} 
\begin{subfigure}[t]{0.3\textwidth}
\includegraphics[width=\textwidth]{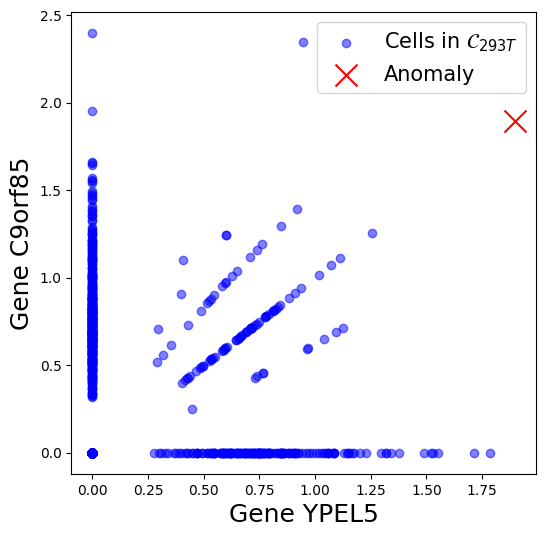}
\label{fig:local_1}
\caption{$\textbf{a}^{(10)}$ in \(\mathcal{C}_{293T}\).}
\end{subfigure}
\hfill
\begin{subfigure}[t]{0.3\textwidth}
\includegraphics[width=\textwidth]{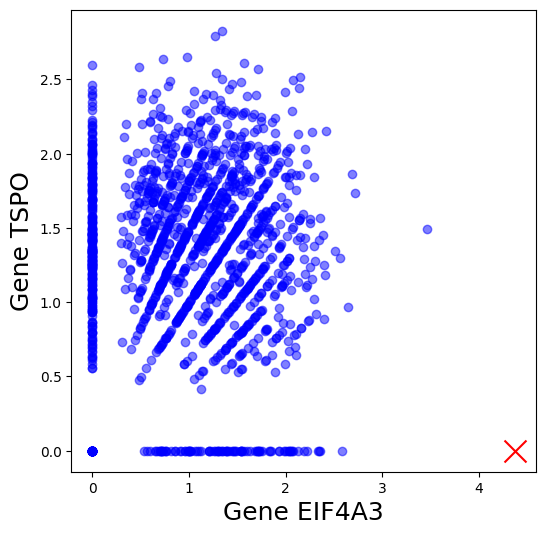}
\label{fig:local_2}
\caption{$\textbf{a}^{(35)}$ in \(\mathcal{C}_{293T}\).}
\end{subfigure}
\hfill
\begin{subfigure}[t]{0.3\textwidth}
\includegraphics[width=\textwidth]{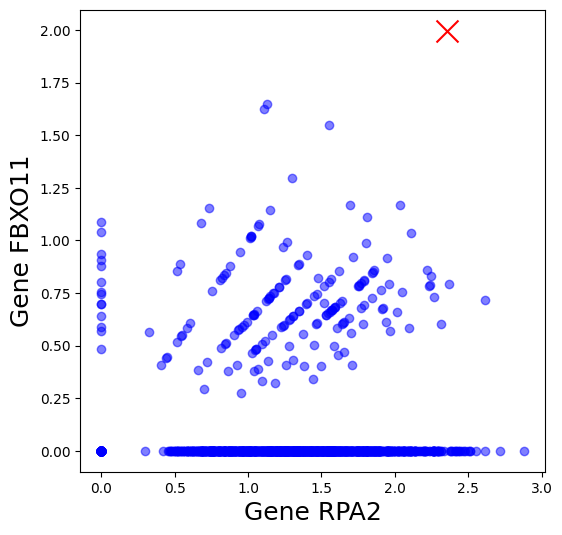}
\label{fig:local_3}
\caption{$\textbf{a}^{(70)}$ in \(\mathcal{C}_{293T}\).}
\end{subfigure}

\caption{SiNNe-derived explanations for typical anomalies sampled from $\mathcal{A}_0$.
Each subfigure demonstrates a unique gene subspace that distinguishes the anomaly (``\textcolor{red}{x}") from the main 293T cell population (blue dots), confirming the heterogeneity of the zero-score group.}
\label{fig:multi_scale}
\end{figure}

The results, presented in \autoref{fig:multi_scale}(a)-(c), strongly support our hypothesis.
Despite their identical IDK scores and proximity in the t-SNE plot, each of the three anomalies was separated from $\mathcal{C}_{293T}$ by a unique subspace of genes:
\begin{itemize}
\item $\textbf{a}^{(10)}$(I\#626) is distinguished by its expression of \textit{YPEL5} and \textit{C9orf85}.
\item $\textbf{a}^{(35)}$(I\#712) is distinguished by \textit{EIF4A3} and \textit{TSPO}.
\item$\textbf{a}^{(70)}$(I\#1308) is distinguished by \textit{RPA2} and \textit{FBXO11}.
\end{itemize}

The fact, that not only the top-3 anomalies but also other members sampled from the broader zero-score anomaly group are explained by different gene subspaces, provides compelling evidence for the heterogeneity hypothesis.
It confirms that $\mathcal{A}_0$ is not a monolithic cluster but rather a ``catch-all bin" for various types of biologically distinct anomalies that IDK successfully detects.

This finding powerfully supports our central argument: normal points are similar, but anomalies are not.
Similar scores of anomalies do not necessarily imply similar biological characteristics.
While the IDK score identifies these cells as potential anomalies, it cannot elucidate the diverse biological mechanisms driving their anomalous states.
Our analysis reveals that this group of top-ranking anomalies is in fact biologically heterogeneous, with each cell distinguished by a unique gene subspace.
This highlights a critical limitation of relying solely on anomaly scores:
an anomaly score tells us that a cell is anomalous, but only an instance-level explanation can reveal why.

\subsubsection{Explaining Complex Anomalies in Diverse Scenarios}

While the Tutorial dataset illustrated straightforward  anomalies in each cluster of one type of normal cell, real-world data often presents greater complexity. 
In this subsection, we show how our framework adapts to two common and challenging scenarios: (1) explaining a ``cross-type" anomaly, where an anomaly's labeled cell type differs from that of its nearest normal neighbors, and (2) identifying and explaining meaningful anomalies within a single large dataset with high cellular diversity. 

\paragraph{Case 1: Dual-Reference Explanation for Cross-Type Anomalies}

A more complex scenario is the presence of ``cross-type" anomaly: the cell type of an anomaly is different from that of the majority of its 10 nearest normal cells.
This  suggests an ambiguous state of the anomaly.
Our framework can explain this with a dual-reference explanation.

We use the Airway dataset as an example.
The top-ranked anomaly $\textbf{a}^{(1)}$ (I\#1601) is labeled ``Basal", but its nearest normal neighbors all belong to the ``Club" cell type.
To explain it, we generate two distinct explanations for the anomaly: one comparing it against the ``Basal" cell population and the other against the ``Club" cell population.

\begin{figure}[ht]
  \centering
  \begin{subfigure}[t]{0.32\textwidth}
    \includegraphics[width=\textwidth]{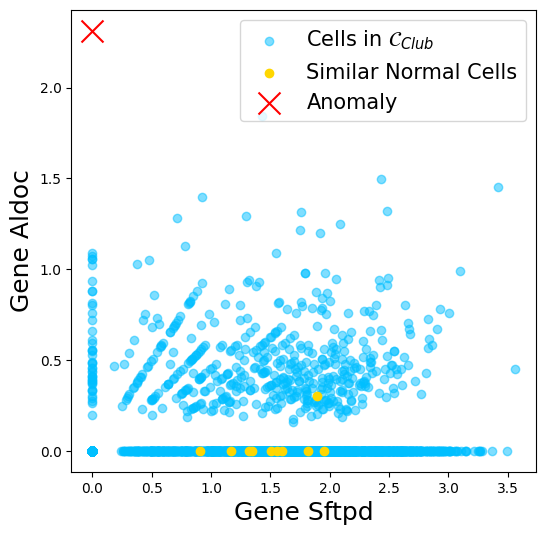}
    \caption{$\textbf{a}^{(1)}$ in $\mathcal{C}_{Club}$}
    \label{fig:1601Basal_in_club}
  \end{subfigure}
  \hfill
  \begin{subfigure}[t]{0.32\textwidth}
    \includegraphics[width=\textwidth]{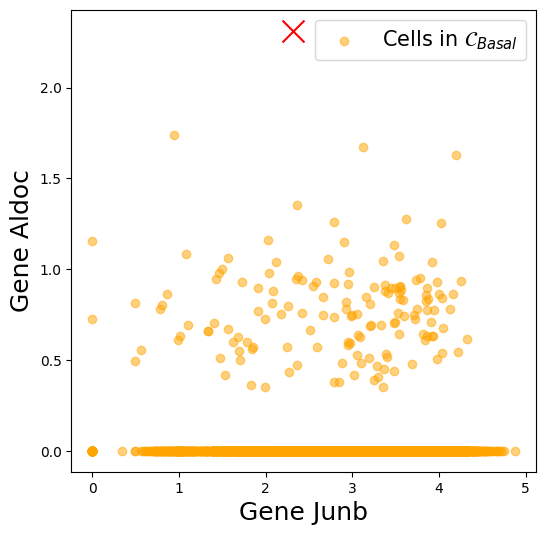}
    \caption{$\textbf{a}^{(1)}$ in $\mathcal{C}_{Basal}$}
    \label{fig:1601Basal_in_basal}
  \end{subfigure}
  \hfill
  \begin{subfigure}[t]{0.32\textwidth}
    \includegraphics[width=\textwidth]{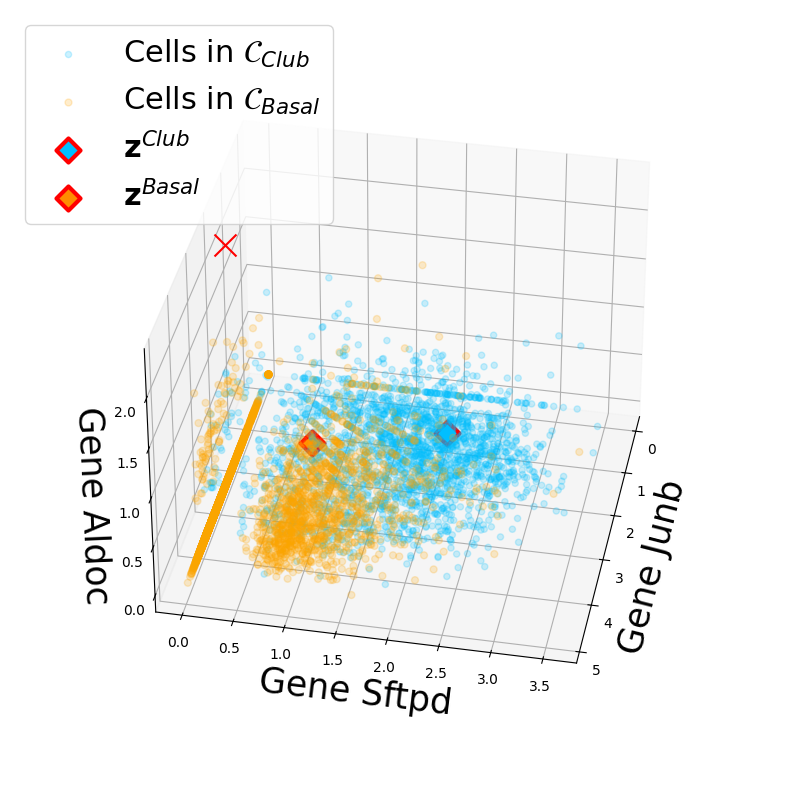}
    \caption{$\textbf{a}^{(1)}$ in $\mathcal{C}_{Basal}$ and $\mathcal{C}_{Club}$}
    \label{fig:1601Basal_in_basal_and_club}
  \end{subfigure}
  \caption{Dual-reference explanation for the cross-type anomaly $\textbf{a}^{(1)}$ in the Airway dataset. The visualizations show its clear separation from (a) its nearest ``Club" neighbors, (b) its own ``Basal" population, and (c) both populations simultaneously.}
  \label{fig:airway_anomaly_explanation_merged}
\end{figure}

As shown in \autoref{fig:airway_anomaly_explanation_merged}, SiNNe identifies distinct gene subspaces that separate $\textbf{a}^{(1)}$ from each of the Club and Basal populations, as well as from a combined set.
The combined view (\autoref{fig:1601Basal_in_basal_and_club}) demonstrates its uniqueness, showing that it is anomalous with respect to both of its related cell types.

\paragraph{Case 2: Pre-filtering for Highly Complex Datasets}

When analyzing large datasets with numerous cell types, the detection can be dominated by cells from globally rare populations.
They are ``trivially rare"\footnote{When IDK was applied to the original, unfiltered Crohn dataset, the top-50 anomalies were scattered across 13 different cell types. 
Notably, 40\% of these top anomalies (20 out of 50) belonged to minor populations that were subsequently excluded due to low abundance (each $<$10\% of the data), including some of the rarest types like ``Group3 ILC'', ``Fibroblasts'', and ``IgM plasma cells''.} and may mask more subtle but biologically significant anomalies within the major cell types.

\begin{figure}[ht]
    \centering
    \includegraphics[width=0.4\textwidth]{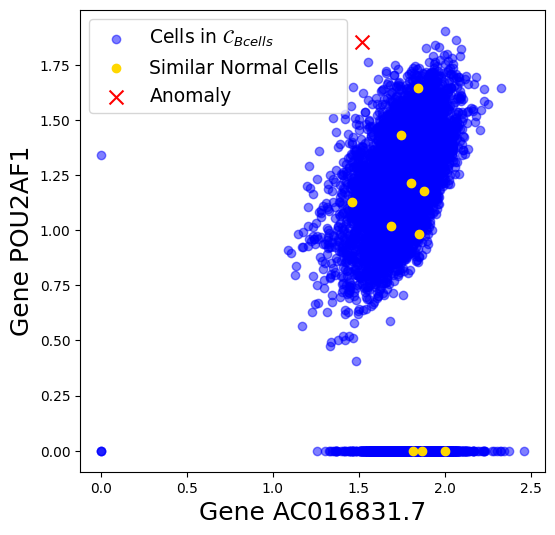}
    \caption{An intra-type anomaly $\textbf{a}^{(1)}$ identified in the Crohn dataset \textit{after} filtering out minor cell populations. The visualization is in the gene subspace of \emph{AC016831.7} and \emph{POU2AF1}.}
    \label{fig:crohn_filtered_example}
\end{figure}

To avoid this, we advocate for a pre-filtering strategy. For instance, in the Crohn’s disease dataset containing 27 annotated cell types, we first filtered out all populations that each constituted less than 10\% of the total data. 
This strategic removal allows the analysis to focus on well-represented populations (in this case, B cells, TRM, and IgA plasma cells) and uncovers anomalies that would otherwise be overlooked.
After filtering, the framework can identify subtle anomalies within these major groups.
For example, the top-ranked anomaly detected was an  anomaly in the B cells population, as shown in \autoref{fig:crohn_filtered_example}.

This highlights how filtering enables the discovery of meaningful, non-obvious anomalies within large cell populations.
The filtering strategy is essential for applying our framework effectively to large and complex biological datasets.

\textbf{Subsection Summary}. The explanations provided by the framework enables us to elucidate the following insights:
\begin{itemize}
    \item Do not be misled by a visualization method such as t-SNE which may show that a group of anomalies is in the same region of t-SNE's transformed space. Despite this apparent similarity, our framework reveals that each anomaly can be explained using a distinct gene subspace, even though they may have the same anomaly score. 
    \item Beware of hidden anomalies that may be masked by a group of `easy' anomalies in a dataset. A simple filter shall be applied in order to examine the existence of these hidden anomalies. The framework can provide the explanation for each and every one of these anomalies.
    \item For each anomaly, the framework provides two types of explanations via (i) a gene subspace which differentiates the anomaly from a group of normal cell which have some members closest to the anomaly; and (ii) a gene subspace which characterizes the group of normal cells. We show that, in a complex dataset such as the Airway dataset, there may exist more than one group of normal cells that are related to the anomaly.
\end{itemize}

\subsection{Quantitative Comparison of Anomaly Detectors}
\label{sec:comparison detectors}
This subsection addresses the fourth question. To quantitatively evaluate the quality and consistency of anomalies detected by different detectors, we adopt the evaluation method proposed by Ting et al.~\cite{Ting2025}.
Their work introduces a method to assess the consistency of self-verifiable detectors\footnote{There are two different types of detectors, \textit{i.e.}, self-verifiable and self-unverifiable detectors~\cite{Ting2025}. The former can verify whether the detected anomalies are different from the normal instances and rare in the given dataset without ground-truths, but the latter cannot. This method is applicable to self-verifiable detectors only, and it is meaningless to use self-unverifiable detectors because the detected ``anomalies'' cannot be understood as anomalies (see \cite{Ting2025} for details).} without ground truths.
The method defines a score, denoted as A$|$B, where A and B denote the anomaly scores computed from detectors A and B.  
The A$|$B score measures the minimum proportion of detector A's ranked list that must be examined in order to retrieve all of the top \(e\) anomalies identified by detector B\footnote{The A$|$B score is functionally identical to the $\Gamma(\text{B}|\text{A})$ metric defined in Methods section, with the parameter $e$ here same to the original definition. The key distinction is that the metric is now used to compare different anomaly \textbf{detectors} (Detector A vs. B), whereas previously it was used to compare \textbf{feature sets} (e.g., HVG vs. PCA).}. The results of applying this pairwise comparison to our four datasets are summarized in \autoref{tab:detector_comparison}.

\begin{table}[ht]
  \centering
  \caption{Quantitative comparison of anomaly detection consistency across different methods and datasets.} 
  \label{tab:detector_comparison}
  \sisetup{table-format=1.2} 
  \begin{tabular}{lSSSS}
    \toprule
    \textbf{Dataset} & {\textbf{LOF$|$IDK}} & {\textbf{IDK$|$LOF}} & {\textbf{IForest$|$IDK}} & {\textbf{IDK$|$IForest}} \\
    \midrule
    Tutorial & 0.47 & 0.24 & 0.46 & 0.32 \\
    \multicolumn{5}{l}{\footnotesize \textit{\quad Comment:} $\text{LOF}|\text{IDK} \gg \text{IDK}|\text{LOF} \approx 0.24$; \quad $\text{IForest}|\text{IDK} > \text{IDK}|\text{IForest} \approx 0.32$} \\
    \addlinespace 
\midrule
    Airway   & 0.97 & 0.85 & 0.99 & 0.86 \\
    \multicolumn{5}{l}{\footnotesize \textit{\quad Comment:} $\text{LOF}|\text{IDK} > \text{IDK}|\text{LOF} > 0.8$; \quad $\text{IForest}|\text{IDK} > \text{IDK}|\text{IForest} > 0.8$} \\
    \addlinespace
\midrule
    Tonsil   & 0.54 & 0.43 & 0.72 & 0.77 \\
    \multicolumn{5}{l}{\footnotesize \textit{\quad Comment:} $\text{LOF}|\text{IDK} > \text{IDK}|\text{LOF} \approx 0.43$; \quad $\text{IForest}|\text{IDK} \approx \text{IDK}|\text{IForest} \approx 0.7$} \\
    \addlinespace
\midrule
    Crohn    & 0.76 & 0.23 & 0.77 & 0.42 \\
    \multicolumn{5}{l}{\footnotesize \textit{\quad Comment:} $\text{LOF}|\text{IDK} \gg \text{IDK}|\text{LOF} \approx 0.23$; \quad $\text{IForest}|\text{IDK} \gg \text{IDK}|\text{IForest} \approx 0.42$} \\
    \bottomrule
  \end{tabular}
\end{table}

The result reveal several key insights.
On the Airway dataset, both LOF (LOF$|$IDK = 0.97) and IForest (IForest$|$IDK = 0.99) demonstrate poor agreement when tasked with identifying anomalies previously detected by IDK. High scores indicate that the top anomalies identified by IDK are not highly ranked by LOF or IForest.
Conversely, IDK's performance in finding anomalies from LOF and IForest is also limited, though comparatively better (scores of 0.85 and 0.86).

The Crohn and Tutorial datasets highlight a significant asymmetry.
For example, on the Crohn dataset, IDK is highly effective in recovering the anomalies found by LOF (IDK$|$LOF score of 0.23), whereas LOF is much less effective in finding IDK's anomalies (LOF$|$IDK score of 0.76).
This suggests that in these biological contexts, the set of top anomalies identified by LOF may be a near-subset of those found by IDK.
A similar, though less pronounced, asymmetry is seen between IDK and IForest.

In contrast, the Tonsil dataset shows a more symmetric case, level of agreement between IForest and IDK (0.72 for IForest$|$IDK vs. 0.77 for IDK$|$IForest), suggesting that neither method's anomaly list is a clear subset of the other's.

\subsubsection{Do Different Detectors Capture Different Anomalies?}

The above quantitative analysis revealed that while IDK is often superior (having IDK$|$Detector scores lower than Detector$|$IDK scores in almost all comparisons), disagreements between detectors are common.
This leads to a crucial part of our research question: do these disagreements reflect that detectors capture different, biologically meaningful anomalies? 

To answer this, we designed a qualitative experiment to directly visualize the detection capability of each detector. The procedure is as follows:

\begin{enumerate}
    \item We select a high-ranking anomaly identified by one detector (\textit{e.g.}, a top ranked anomaly from IDK).
    \item We then check the rank of this specific anomaly in the lists produced by the other detectors (LOF and IForest). An anomaly that is ranked highly by one detector but has a very low rank (\textit{i.e.}, it is considered normal) by others would serve as a strong evidence of detector-specific discovery.
    \item Finally, we use SiNNe to find the 2D or 3D gene subspace that best explains the anomalous status of this chosen anomaly and visualize its position relative to all other cells. This allows us to understand the biological characteristics of a detector-unique anomaly.
\end{enumerate}

\begin{figure}[ht]
\centering
\begin{subfigure}{0.32\textwidth}
\centering
\includegraphics[width=\textwidth]{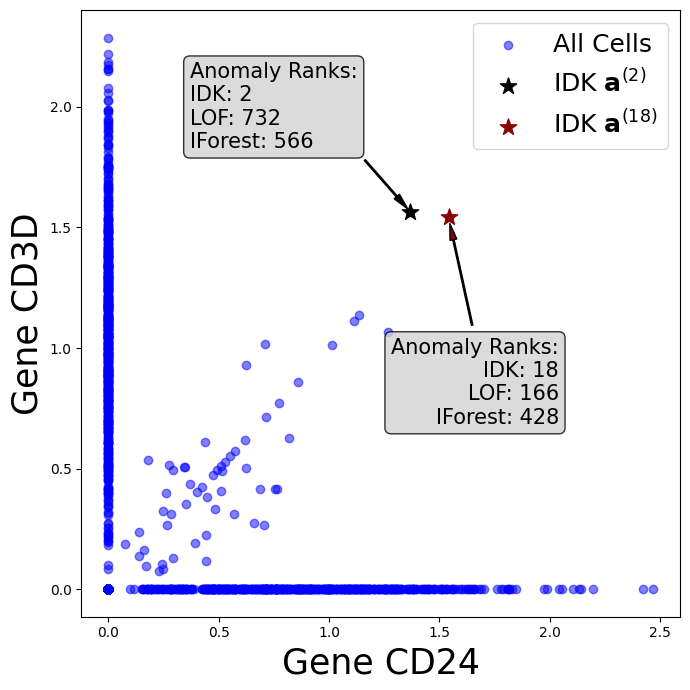}
\caption{IDK-specific $\textbf{a}^{(2)}$ and $\textbf{a}^{(18)}$}
\label{fig:sub_idk_unique}
\end{subfigure}
\hfill
\begin{subfigure}{0.32\textwidth}
\centering
\includegraphics[width=\textwidth]{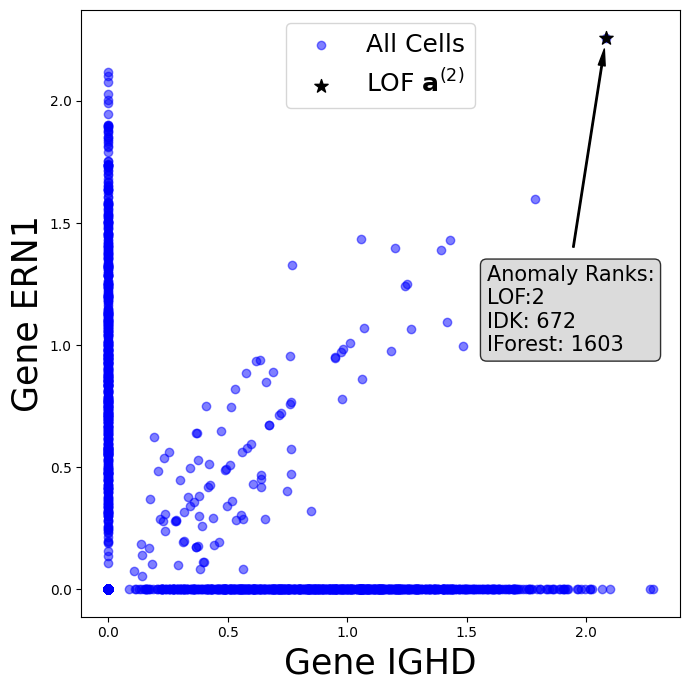}
\caption{LOF-specific $\textbf{a}^{(2)}$}
\label{fig:sub_lof_unique}
\end{subfigure}
\hfill
\begin{subfigure}{0.32\textwidth}
\centering
\includegraphics[width=\textwidth]{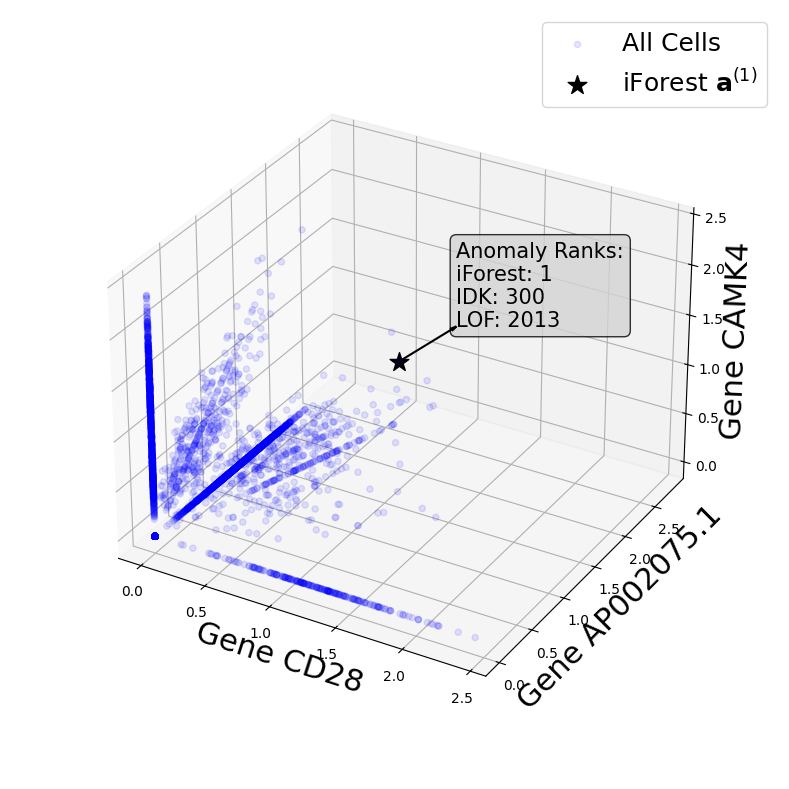}
\caption{IForest-specific $\textbf{a}^{(1)}$}
\label{fig:sub_iforest_unique}
\end{subfigure}
\caption{Visualization of detector-specific anomalies in the Tonsil dataset. Each panel highlights an anomaly (or anomalies) ranked high by one detector but low by the others. The axes represent the explanatory gene subspaces found by SiNNe. The rank discrepancies provide clear evidence of different detection capabilities. (a) IDK $\textbf{a}^{(2)}$ and $\textbf{a}^{(18)}$ are top-ranked by IDK but not by LOF or IForest. (b) LOF $\textbf{a}^{(2)}$ is a top LOF anomaly, but is missed by IDK and IForest. (c) IForest $\textbf{a}^{(1)}$ is the top IForest anomaly but has low ranks in IDK and LOF. }
\label{fig:detector_qualitative_comparison_combined}
\end{figure}

The Tonsil dataset is used as an example, where the A$|$B scores suggested significant yet symmetric disagreement.
\autoref{fig:detector_qualitative_comparison_combined} illustrates three cases where anomalies highly ranked by one detector are missed by the others.
Each panel illustrates an anomaly that receives a high rank from one detector but a low rank from the other two.
The explanatory gene subspace for each case is generated by SiNNe to highlight the specific anomaly.

\begin{itemize}
    \item \textbf{IDK's Unique Detections (\autoref{fig:sub_idk_unique}):}
    IDK's high ranking of anomalies $\textbf{a}^{(2)}$ and $\textbf{a}^{(18)}$ are explained by genes \textit{CD24}
    and \textit{CD3D}, while both of the two high ranking anomalie are seen much less anomalous by LOF and IForest.
    
    \item \textbf{LOF's Unique Detections (\autoref{fig:sub_lof_unique}):}  
    The LOF-Unique anomaly is ranked 2nd by LOF but only 672-th by IDK and 1603-th by IForest.
    Its anomalous status is explained by genes \textit{IGHD} and \textit{ERN1}.

    \item \textbf{IForest's Unique Detections (\autoref{fig:sub_iforest_unique}):} The IForest-Unique anomaly is ranked 1st by IForest, it is overlooked by IDK (rank 300) and especially by LOF (rank 2013).
    The visualization in the subspace of \textit{CD28}, \textit{AP002075.1}, and \textit{CAMK4} illustrates its anomalous status.
\end{itemize}

The results clearly show that the definition of an ``anomaly" is detector-dependent.
What one algorithm identifies as a top anomaly can be considered as normal by another, indicating that these detectors are not replaceable.

This divergence is consistent with the differences in their algorithmic biases when applied to high-dimensional transcriptomics data.
IDK excels at identifying complex distributions, while LOF is specialized for detecting anomalies based on local density differentials.
In contrast, IForest is good at finding points that are easily separable through partitioning.
This visual analysis provides an interpretation for the quantitative results from \autoref{tab:detector_comparison}: the symmetric disagreement on the Tonsil dataset is indeed a scenario of complementarity, where each detector captures biologically distinct anomalies.

In summary, this analysis provides a multi-faceted answer to the fourth question.
Our quantitative results show that IDK is generally the most comprehensive detector, making it an excellent primary choice.
However, the visual evidence demonstrates that no single detector can guarantee to identify every anomaly in a dataset.
For a certain distribution, multiple detectors are required to identify anomalies that are overlooked by an individual detector. 
This approach allows researchers to move beyond a single definition of an ``anomaly" and capture a richer, more complete landscape of rare cellular states.

\subsubsection{Self-Unverifiability of Deep Learning Detectors: A Case Study with Autoencoder}
\label{sec:dl_detector_verifiability} 

To address the question as to why deep learning (DL) detectors were not included in our primary comparative analysis, we present a case study using a standard Autoencoder (AE) \cite{chen2018autoencoder}, a representative DL-based anomaly detector \cite{pang2021deep}. 
We argue that while powerful, such models are typically not self-verifiable, i.e., it does not have a necessary property in our explanatory framework.

An AE is trained to minimize reconstruction error, effectively learning a latent representation of ``normal" data. 
Cells that deviate from this learned norm are poorly reconstructed, resulting in a high error score that marks them as anomalies. A crucial difference, however, emerges when interpreting these scores. The high reconstruction error for an anomaly is a scalar value resulting from a complex, non-linear transformation through the AE's opaque layers. The score itself does not convey a verifiable property of the cell's relationship to the data; it is merely the output of a black-box model.

To quantitatively assess this lack of verifiability, we adopt the methodology from Ting et al. \cite{Ting2025}.
A key indicator of a self-verifiable detector is a distinct gap between the scores of genuine anomalies and the scores of normal instances. 
Conversely, a non-verifiable detector tends to produce a smooth continuum of scores where such a gap is absent.

To measure this, we use the sorted anomaly scores to define a reference range based on the first 100 normal cells.
In \autoref{tab:ae_verifiability}, we denote this setting as ``$i$th Most Normal" with $i=100$.
We then identify how many of the top-ranked anomalous cells have scores that fall within this same reference range.
The rank of the last anomalous cell to do so is denoted as \textbf{$\hat{j}$}.
We expect a very small $\hat{j}$ (i.e., $\hat{j} \ll 100$) if the anomalies are rare and different from the normal.

\begin{table}[ht]
\centering
\caption{Self-verifiability analysis of the Autoencoder (AE) detector \cite{chen2018autoencoder} versus a self-verifiable detector (IDK) \cite{Xu2020IDK}. The ``$\hat{j}$th Most Anomalous" rank indicates how many top-ranked cells have scores nearly identical to the single most anomalous cell. Large values are characteristic of self-unverifiable detectors.}
\label{tab:ae_verifiability}
\label{tab:ae_vs_idk_verifiability}
\begin{tabular}{l|c|cc}
\toprule
\multirow{2}{*}{\textbf{Dataset}} & \multirow{2}{*}{\textbf{\tabincell{c}{$i$th Most Normal}}} & \multicolumn{2}{c}{\textbf{$\hat{j}$th Most Anomalous}} \\
\cmidrule(lr){3-4}
& & \multicolumn{1}{c}{\textbf{AE}} & \multicolumn{1}{c}{\textbf{IDK}} \\
\midrule
Tonsil   & 100 & 3\,948 & 6 \\
Crohn    & 100 & 9\,721 & 26 \\
\bottomrule
\end{tabular}
\end{table}

We applied this analysis to the anomaly scores from both the AE and IDK in two datasets as an example (\autoref{tab:ae_verifiability}). 
The large $\hat{j}$ values of AE in \autoref{tab:ae_verifiability} suggests that there are significantly more `anomalies' which are similar to each other than normal cells! In contrast, IDK has the number of anomalies which is significantly smaller than normal cells. This shows that the anomalies detected by IDK are rare.


This is the characteristic of the Data-Learned Parameter of an
Assumed Model (LPaM) detector \cite{Ting2025} such as AE. As argued therein, the onus is on each LPaM detector to provide its own mechanism for verifying rarity. This inherent deficiency makes an LPaM detector incompatible with our framework's need for verifiable inputs.


\subsection{A Multi-faceted Validation of scCAD's Anomaly Groups}
\label{sec:scCAD}
A key challenge in rare cell detection is not only identifying candidate anomaly clusters but also rigorously validating their biological significance.
While methods such as scCAD \cite{Xu2024scCAD} are effective at detecting small cell populations, their group-based outputs necessitate further investigation.
This subsection is to answer the last question.

We first establish a formal definition for what constitutes a subspace anomaly group:

\begin{definition}[Subspace Anomaly Group]
\label{def:subspace_anomaly_group}
A \textbf{subspace anomaly group} is a small, rare group of cells that satisfies two key properties in some subspaces of genes:
\begin{enumerate}
    \item \textbf{Collective Distinctiveness:} The anomaly group must be demonstrably distinct from the majority of normal groups in a subspace.
    \item \textbf{Individual Separability:}  Every individual member of the group is clearly separated from its nearest normal cells in a subspace. 
\end{enumerate}
\end{definition}

Note that the subspaces in the two properties above are likely to be different. 

We perform validations with this definition as a guide, starting with a quantitative assessment and progressing to detailed visual confirmation.

\subsubsection{Quantitative Validation of Anomaly Group via $\textit{IDK}^{2}$.}
Before examining the properties of the anomaly groups detected by scCAD, we first need to verify whether they are truly anomalous.
To do this, we performed a validation using the two level Isolation Distributional Kernel ($\textit{IDK}^{2}$) for group anomalies \cite{ting2023isolation}.
For the Airway, Tutorial, and Tonsil datasets, we first identified all anomaly groups detected by scCAD.
Next, for each anomaly group, we generated normal groups from all clusters that were not identified as anomalous by the algorithm and each normal group was constructed by sampling cells from one normal cluster at a time.
Each normal group had the same number of cells as the anomaly group.
This process was repeated for every available normal cluster.

Once a set of groups (with ground-truth normal and anomaly groups identified by scCAD) was established, we used $\textit{IDK}^2$ to compute a distributional anomaly score for every group and assess its detection performance in terms of the Area Under the Curve (AUC), based on the ground-truth labels.

The results were unequivocal.
The anomaly groups detected by scCAD were almost perfectly separable from their normal groups, achieving an AUC of \textbf{0.998} on the Airway dataset, and \textbf{1.000} on the Tonsil and Tutorial datasets, providing a quantitative verification that the groups identified by scCAD are valid anomalies.

\begin{table}[ht]
  \centering
       \caption{
    Summary of the scCAD-identified anomaly groups. Group Rank is determined by $\textit{IDK}^2$. The ``Avg. Normalized Manhattan Distance" quantifies the separation from the parent cluster in the explanatory subspace. 
    The distance is computed as follows: a mean distance is calculated for each individual anomaly cell, defined as the mean of the Manhattan distances to its $k$-nearest normal neighbors within the explanatory subspace, normalized by the number of dimensions.
    The overall distance for an anomaly group is then computed by averaging these individual mean distances across all cells in the group.
    A higher distance indicates a clearer separation.
  }
  \begin{tabular}{@{}l|c|c|c@{}}
    \toprule
    \textbf{Dataset} & \tabincell{c}{\textbf{\# Anomaly Groups} \\ \textbf{Detected}} & \tabincell{c}{\textbf{Anomaly Group Rank} \\ \small{(Lower is more anomalous)}} & \tabincell{c}{\textbf{Avg. Normalized} \\ \textbf{Manhattan Distance}} \\
    \midrule
    \multirow{2}{*}{Airway} & \multirow{2}{*}{4} & First  & 1.03 (\autoref{fig:airway_g1}) \\
                     &                    & 5-th   & 0.82 (\autoref{fig:airway_g3}) \\
    \midrule
    \multirow{2}{*}{Tonsil} & \multirow{2}{*}{5} & First  & 0.36 (\autoref{fig:tonsil_g1})  \\ 
                     &                   & 5-th   & 0.22 (\autoref{fig:tonsil_g3}) \\
    \midrule
    Tutorial & 1                  & First  & 0.94 (\autoref{fig:tutorial_g1}) \\
    \bottomrule
  \end{tabular}
    \label{tab:sccad_summary}
\end{table}

\begin{figure}[t!]
  \centering
  \begin{subfigure}{0.48\textwidth}
\includegraphics[width=\linewidth]{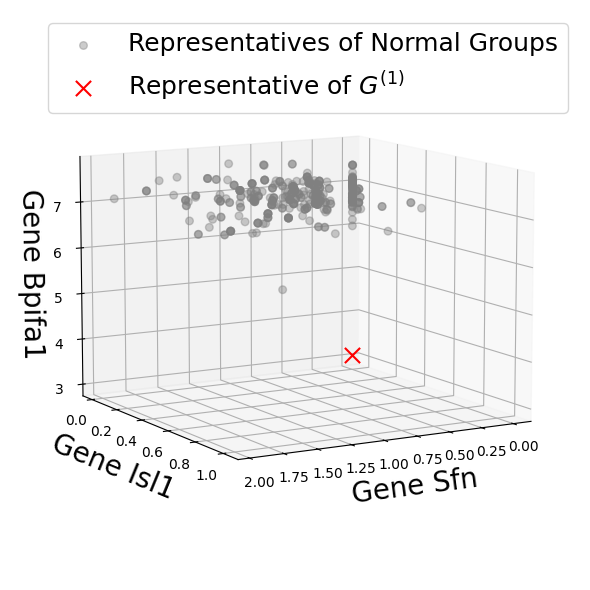}
    \caption{Airway $G^{(1)}$ }
    \label{fig:rep_g1_airway}
  \end{subfigure}
  \hfill
  \begin{subfigure}{0.48\textwidth}
\includegraphics[width=\linewidth]{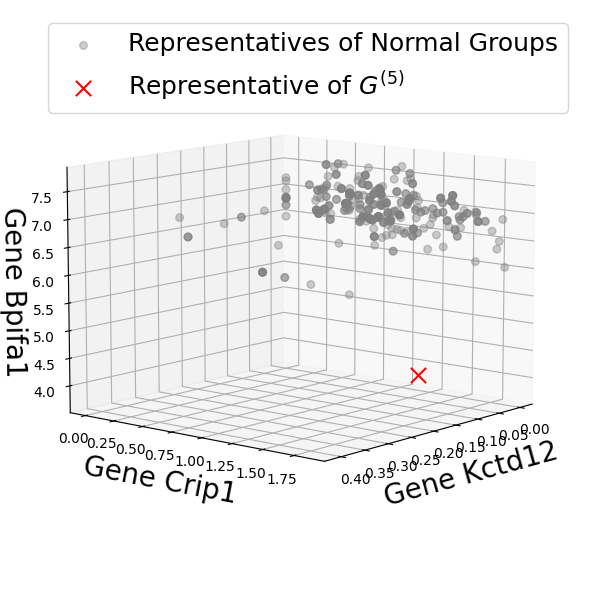}
    \caption{Airway $G^{(5)}$}
    \label{fig:rep_g5_airway}
  \end{subfigure}
   \vspace{0.5cm} 
   
   \begin{subfigure}{0.32\textwidth}
\includegraphics[width=\linewidth]{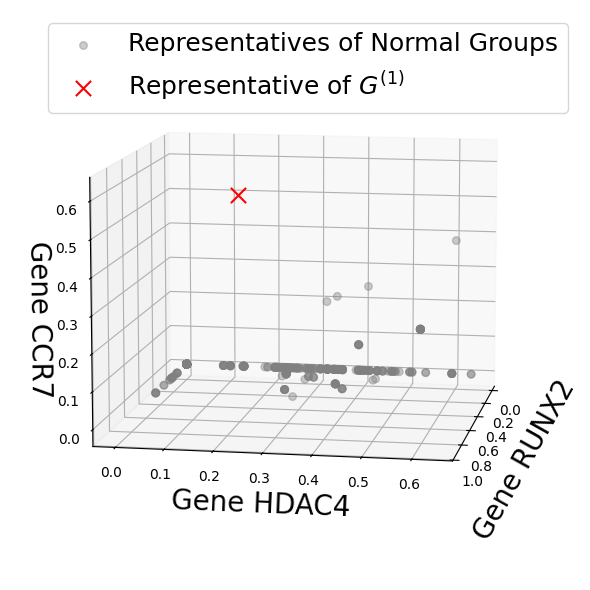}
    \caption{Tonsil $G^{(1)}$ }
    \label{fig:rep_g1_tonsil}
  \end{subfigure}
  \hfill
  \begin{subfigure}{0.32\textwidth}
    \includegraphics[width=\linewidth]{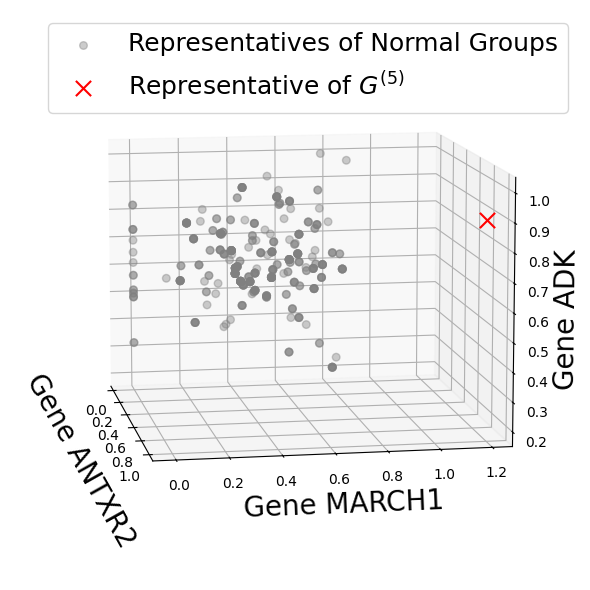}
    \caption{Tonsil $G^{(5)}$}
    \label{fig:rep_g36_tonsil}
  \end{subfigure}
  \hfill
   \begin{subfigure}{0.32\textwidth}
    \includegraphics[width=\linewidth]{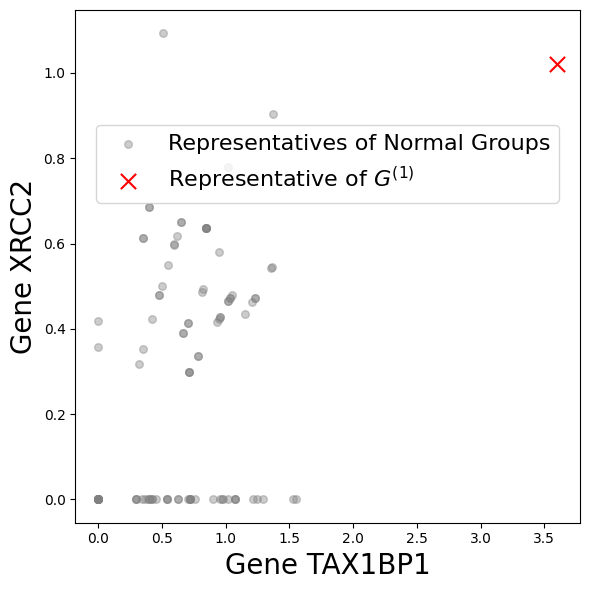}
    \caption{Tutorial $G^{(1)}$}
    \label{fig:rep_g1_tutorial}
  \end{subfigure}
  \caption{Comparison of representatives from anomaly and normal groups in the three datasets. The superscript in $G^{(n)}$ denotes the group's rank provided by $IDK^2$. (\subref{fig:rep_g1_airway}, \subref{fig:rep_g5_airway}) For the Airway dataset, the representatives from both the highest-ranked anomaly group $G^{(1)}$ and the lowest-ranked anomaly group $G^{(5)}$ are clearly separated from normal groups.
(\subref{fig:rep_g1_tonsil}, \subref{fig:rep_g36_tonsil}) A similar clear separation is observed in the Tonsil dataset for both its highest-ranking ($G^{(1)}$) and lowest-ranking ($G^{(5)}$) anomaly groups.
(\subref{fig:rep_g1_tutorial}) The single anomaly group in the Tutorial dataset, $G^{(1)}$, also shows a distinct separation from its normal groups. All these confirms the distinctiveness of the scCAD-identified anomaly groups.}
\label{fig:rep_comparison}
\end{figure}

\begin{figure}[t!]
  \centering
  \begin{subfigure}{0.48\textwidth}
  
    \includegraphics[width=\linewidth]{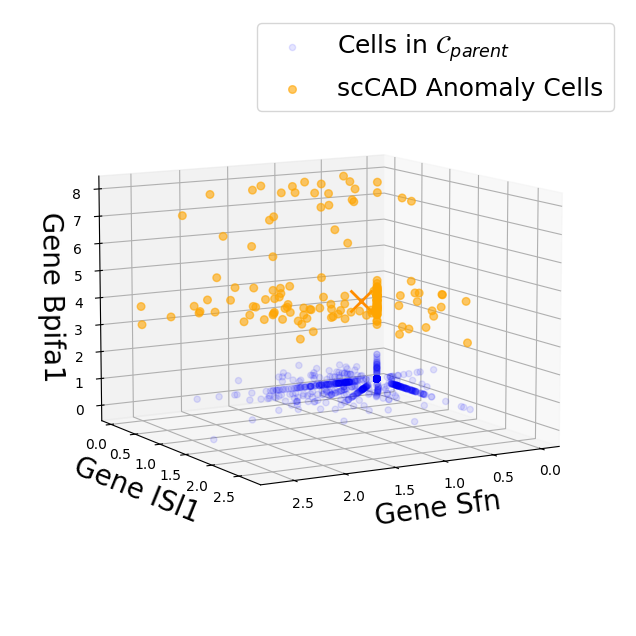}
    \caption{Airway $G^{(1)}$}
    \label{fig:airway_g1}
  \end{subfigure}
  \hfill
  \begin{subfigure}{0.48\textwidth}
    \includegraphics[width=\linewidth]{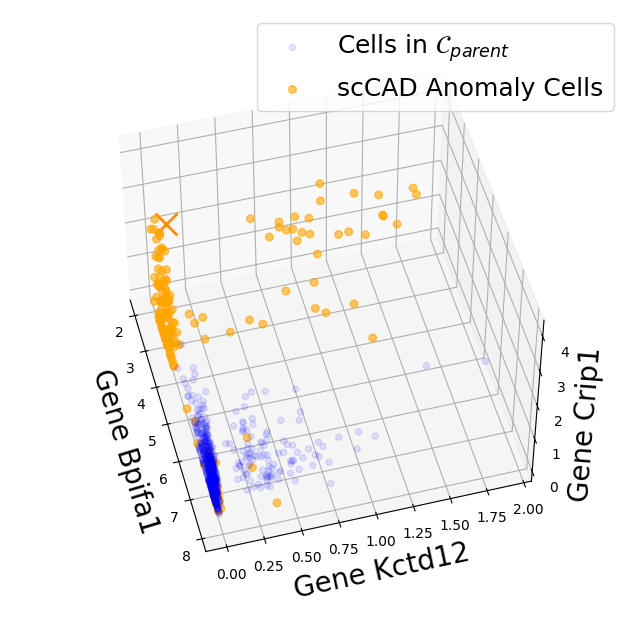} 
    \caption{Airway $G^{(5)}$}
    \label{fig:airway_g3}
  \end{subfigure}
  
  \vspace{0.1cm} 
  
  \begin{subfigure}{0.33\textwidth}
    \includegraphics[width=\linewidth]{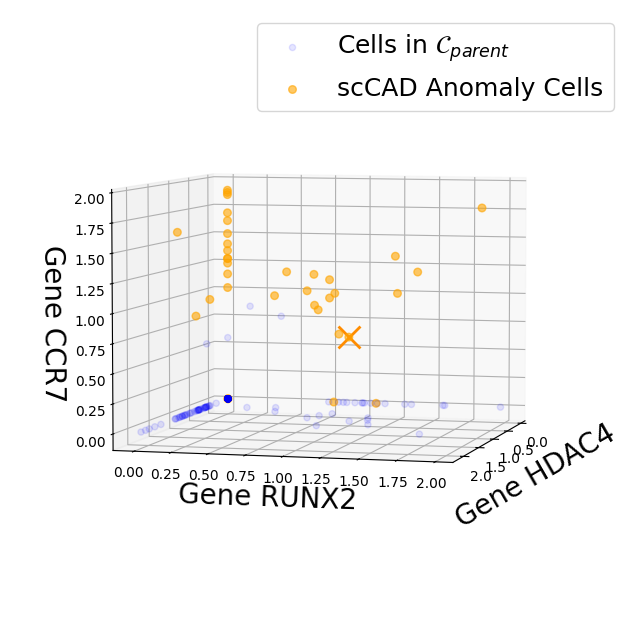}
    \caption{Tonsil $G^{(1)}$}
    \label{fig:tonsil_g1}
  \end{subfigure}
  \hfill
  \begin{subfigure}{0.33\textwidth}
    \includegraphics[width=\linewidth]{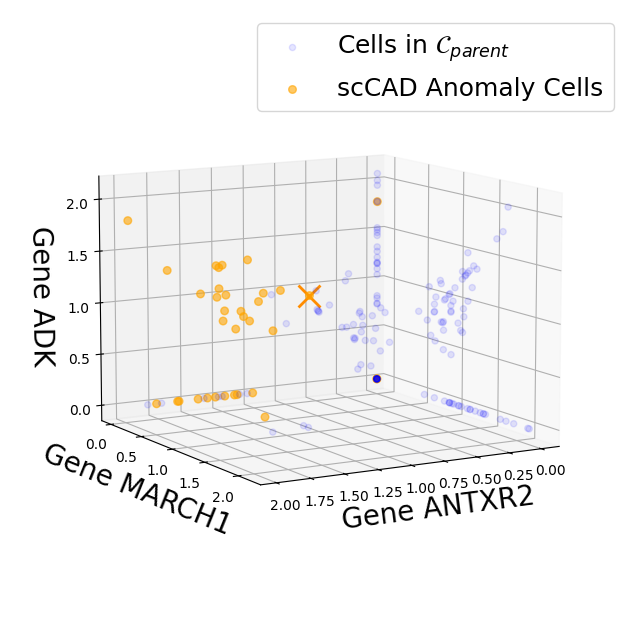} 
    \caption{Tonsil $G^{(5)}$}
    \label{fig:tonsil_g3}
  \end{subfigure}
  \begin{subfigure}{0.3\textwidth}
    \includegraphics[width=\linewidth]{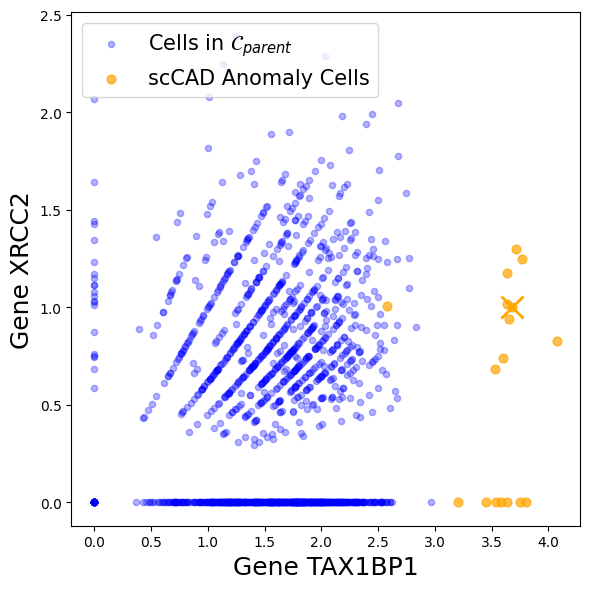}
    \caption{Tutorial $G^{(1)}$}
    \label{fig:tutorial_g1}
  \end{subfigure}
  \caption{Visualizing the scCAD anomaly groups validated by $\textit{IDK}^2$. In Airway,
  (\subref{fig:airway_g1}) the top-ranked anomaly group $G^{(1)}$  is almost perfectly separated from normal groups.
  (\subref{fig:airway_g3}) In contrast, $G^{(5)}$ shows some overlap with its parent cluster. In Tonsil,
  (\subref{fig:tonsil_g1}) and (\subref{fig:tonsil_g3}) shows that both $G^{(1)}$ and $G^{(5)}$ have some overlap with their parent clusters. In Tutorial, (\subref{fig:tutorial_g1}) the top-ranked anomaly group $G^{(1)}$ forms an almost well-separated cluster, with the exception of a single cell that should belong to the normal group.}
  
  \label{fig:idk2_visuals}
\end{figure}

\subsubsection{Assessing the Quality Against the Formal Definition}

As scCAD-detected anomaly groups are verified as truly anomalous groups, to further verify if these groups satisfy the two properties in Definition \ref{def:subspace_anomaly_group}, we applied the two group explanation methods detailed in \autoref{tab:point_vs_group} to the highest- and lowest-ranking anomaly groups.
\begin{enumerate}
    \item First, to test for Collective Distinctiveness, we used $\text{Expl}_{\mathrm{anom}}(\mathbf{r}_G ,R_\mathcal{N})$ to find a subspace that separates the group's representative \(\textbf{r}_G\) from normal representatives \(R_\mathcal{N}\).
    The resulting visualizations are presented in \autoref{fig:rep_comparison}.
    \item Second, to assess Individual Separability, we applied $\text{Expl}_{\mathrm{norm}}(\mathbf{r}_G , G)$ to identify the gene subspace that best characterizes $\mathbf{r}_G$ as a representative of its own group $G$.
    We then used this subspace to visualize all cells of the anomaly group against their parent cluster for inspection, as shown in \autoref{fig:idk2_visuals}.
\end{enumerate}

\autoref{fig:rep_comparison} assesses the Collective Distinctiveness.
The consistent separation between representative anomaly and normal representatives confirms that scCAD groups satisfy this criterion, demonstrating they are distinct.

\autoref{fig:idk2_visuals} assesses the Individual Separability and reveals a more complex picture.
While top-ranked groups (\textit{e.g.}, Airway \(G^{(1)}\)) are well-separated, lower-ranked groups (\textit{e.g.}, Airway \(G^{(5)}\) and Tonsil \(G^{(5)}\)) show considerable overlap with their parent cluster.
This indicates that while the group is collectively identified as anomalous, the property of individual separability might not be universally met, especially in lower-ranked groups.

A direct visual comparison between plots—for instance, between the top-ranked Airway group \(G^{(1)}\) and the lower-ranked \(G^{(5)}\)—can be ambiguous due to differences in their respective gene expression scales.
To provide a standardized metric for this comparison, we calculated the ``Average Normalized Manhattan Distance", with the results and a detailed explanation provided in \autoref{tab:sccad_summary}.
This metric quantitatively confirms our visual assessment: the top-ranked anomaly groups consistently achieve higher distance scores (\textit{e.g.}, 1.03 for Airway \(G^{(1)}\) vs. 0.82 for Airway \(G^{(5)}\)), demonstrating a clearer separation from their parent clusters.
This result also validates that the group ranking provided by $\textit{IDK}^2$ is strongly correlated with the degree of spatial separation in the explanatory gene subspace.
Since scCAD's cluster-decomposition approach is designed to identify anomalous regions rather than perfectly pure groups, such ambiguity at the group boundary is not entirely unexpected.

In summary, we show that scCAD-detected anomaly groups generally satisfy the first property of collective distinctiveness within their subspaces.
However, the analysis at the individual cell level indicates that while the group as a whole is anomalous, some members may exhibit a degree of overlap with the nearest normal cells.
To verify a group anomaly from any detector, we propose a two-stage procedure.
First, a group anomaly detector is used to identify candidate rare cell clusters.
Second, a comprehensive validation stage is performed.
This involves first confirming the group's anomalous status with a quantitative measure (as we demonstrated using $\textit{IDK}^2$), followed by the application of our explanatory framework. 
The framework then serves to identify the gene subspace that makes the group unique and to enable an instance-level inspection of its purity.

\section{Discussion}
A key challenge in single-cell transcriptomics is to explain why a detected rare cell is biologically distinct.
Our proposed framework provides a direct, gene-level explanation as to why a detected cell is rare/anomalous which has the two unique features.
First, we redesigned the typical framework by eliminating the (previously compulsory) PCA transformation step.
This not only preserves the biological meaning that would otherwise lost in the PCA transformation step, but also leads to a better anomaly detection outcome than those reliant on PCA.
Second, our framework integrates the IDK anomaly detector and the significantly improved SiNNe explainer, which are methods well-suited for high-dimensional, sparse data, enabling an effective solution tailored to the challenges of single-cell transcriptomics data.

We tested our framework on several transcriptomics datasets and discovered previously unknown insights on these datasets. The framework was also used to examine the anomaly groups identified by an existing group anomaly detector scCAD.
The explanations provided by the framework uncovered that while scCAD is good at finding ``anomaly groups", it is not perfect, \textit{i.e.},
some detected anomaly clusters may contain normal cells. An additional experiment also shows the role of the framework in validating that multiple detectors must be applied in order to identify all anomalies in some dataset.
All these underscore the need of an explanatory tool for validating single detector of different types (i.e., group anomaly versus point anomaly detectors) as well as multiple detectors of the same type in different scenarios.
By providing a clear, genes-based explanation as to why a cell is anomalous or normal,
our work addresses a critical challenge of inability to provide an explanation for rare cell in the domain of single-cell transcriptomics data.
The framework has the potential in wider applications to provide explainable analyses, that are non-existence at the moment, in other complex biological domains facing similar challenges of data with high dimensionality and high sparsity.

\section{Methods}

The key symbols and notations are provided in \autoref{tab:notation}.

\begin{table}[ht]
\caption{Key symbols and notations. }
\renewcommand{\arraystretch}{1}
\label{tab:notation}
\begin{center}
\begin{tabular}{ cl } 
\toprule
Symbol & Description\\
\midrule
$\mathbf{x}$ & A cell represented as a gene expression vector in $\mathbb{R}^d$.\\

$D$ & Set of $n$ cells $\{\mathbf{x}_i,i=1,\dots,n\}$\\
$\kappa$ &  Isolation Kernel \\
$\phi$ & Feature map of $\kappa$\\

  $\mathcal{K}$ &  Isolation Distributional Kernel \\
  ${\Phi}$ & Feature mean map of $\mathcal{K}$\\
  $t$ & Number of partitionings in Isolation kernel\\
  $H$ & A partitioning of the space\\
  $\theta$ & An isolating partition in partitioning $H$\\
  $\psi$ & Number of isolating partitions in partitioning $H$\\
  $\mathbb{H}_\psi(D)$ & Set of all possible partitionings $H$ derived from $D$\\
  $\textbf{a}$ & An anomalous cell \\
  $\textbf{z}$ & A typical normal cell
  \\
  $G$ & A group anomaly (anomalous group)\\
  $\mathcal{N}$ & A set of normal groups\\
  $R_\mathcal{N}$ & The set of representative cells from the normal groups\\
  $\textbf{r}_G$ & A representative cell of a group anomaly $G$. \\
& (The cell closest to the group's geometric centroid.)\\
  \bottomrule
\end{tabular}
\end{center}
\end{table}
\subsection{Datasets}

We evaluate our framework on four publicly available single-cell transcriptomics datasets, selected to represent a diverse range of biological systems, sample sizes, and complexities:
\begin{itemize}
  \item \textbf{Tutorial} – A synthetic/imputed benchmark with 2 known cell types, 1\,556 cells, and 32\,738 genes, commonly used for method validation.
  \item \textbf{Airway} – Human airway epithelial cells comprising 7 distinct cell types, 7\,193 cells, and 27\,716 genes.
  \item \textbf{Tonsil} – Human tonsil immune microenvironment with 13 annotated cell types, 5\,778 cells, and 36\,601 genes.
  \item \textbf{Crohn} – Intestinal biopsy spatial transcriptomics from Crohn’s disease patients, containing 27 histologically defined regions, 39\,563 spots, and 33\,660 genes.
\end{itemize}

\begin{table}[t]
  \centering
  \caption{Dataset used in our experiments. All subsequent analyses were performed on the top 2,000 highly variable genes (HVGs).}
  \label{tab:datasets}
  \begin{tabular}{l|c|c|c}
    \toprule
    \textbf{Dataset} & \(\mathbf{\#}\)\textbf{Cell Types} & \textbf{\#Cells / \#Spots} & \textbf{\#Genes} \\
    \midrule
    Tutorial & 2  & 1\,556  & 32\,738 \\
    Airway   & 7  & 7\,193  & 27\,716 \\
    Tonsil   & 13 & 5\,778  & 36\,601 \\
    Crohn    & 27 & 39\,563 & 33\,660 \\
    \bottomrule
  \end{tabular}
\end{table}

\autoref{tab:datasets} summarizes the key statistics of each dataset.

\subsection{Preprocessing}

Let \(X\in\mathbb{R}^{n\times d}\) be the raw count matrix of the data extracted from a tissue sample, where \(n\) is the number of cells or spots and \(d\) is the number of genes.
Our proposed pipeline consists of the following steps:

\begin{enumerate}
  \item \textbf{Normalization.}  
    Normalize the total counts per cell (or spot) to the population median via  
    \[
      x_{ij}\leftarrow \frac{x_{ij}}{\sum_{k}x_{ik}}\times \mathrm{median}\bigl(\{\sum_{k}x_{ik}\}_{i=1}^n\bigr).
    \]  
  \item \textbf{Log‐transformation.}  
    Stabilize variance by computing  
    \(\ln(x+1)\).

    \item \textbf{Highly Variable Gene Selection.}
    To reduce the dimensionality of the dataset, feature selection was performed to include only the most informative genes. Highly variable genes (HVGs) were identified using the \texttt{highly\_variable\_genes} function from the Scanpy package (\texttt{sc.pp.highly\_variable\_genes(adata, n\_top\_genes=2000)}). This function identifies genes that exhibit high cell-to-cell variation in the data by modeling the mean-variance relationship. We selected the top 2000 most variable genes for downstream analysis.

  \item \textbf{Low-Expression Gene Filtering.}  
    On the normalized and (optionally) scaled matrix, remove any gene \(j\) whose proportion of zero counts exceeds 90\%:  
    \[
      \frac{1}{n}\sum_{i=1}^{n} \mathbb{I}(x_{ij}=0) > 0.9,
    \]  
    thereby eliminating uninformative genes and further reducing sparsity.  
\end{enumerate}

This pipeline follows current best practices  \cite{LueckenSaunders2019, Dries2021Giotto, SpaNorm2025}  for single‐cell transcriptomics preprocessing—median‐based, size‐factor normalization, log‐transform, and stringent feature filtering—to mitigate technical variation and focus downstream analysis on informative genes.  

\subsection{Anomaly Scoring via Isolation Distributional Kernel}
\begin{definition}[Isolation kernel~\cite{ting2018isolation, qin2019nearest}, IK]
    For any two points $\x,\y\in\R^d$, the Isolation kernel of $\x$ and $\y$ is defined to be the expectation taken over the probability distribution on all partitionings $H\in\mathds{H}_\psi(D)$ that both $\x$ and $\y$ fall into the same isolating partition $\theta[\z]\in H$, where $\z\in\D\subset D$, and $\left|\D\right| = \psi$:
    \begin{equation}
    \begin{aligned}
    \label{eq:IK}
        \k(\x,\y\mid D) &= \E_{\mathds{H}_{\psi}(D)}\left[\mathds{1}(\x,\y\in\theta\mid\theta\in H)\right]\\
        &= \frac{1}{t}\sum_{i=1}^{t}\mathds{1}(\x,\y\in\theta\mid\theta\in H_i)\\
        &= \frac{1}{t}\sum_{i=1}^{t}\sum_{\theta\in H_i}\mathds{1}(\x\in\theta)\mathds{1}(\y\in\theta)
    \end{aligned}
    \end{equation}
    where $\k$ is constructed using a finite number of partitionings $H_i, i\in[t]$, each $H_i$ is created via random subsampling a $\D_i\subset D$, and $\theta$ is a shorthand for $\theta[\z]$.
\end{definition}

IK defines a feature mapping $\phi$ that maps $\x\in\R^d$ to feature vectors $\phi(\x)\in\H$, where $\H$ is a high-dimensional feature space.
Re-expressing \autoref{eq:IK} using $\phi$ gives:
\begin{equation}
\label{eq:IKfm}
    \k(\x,\y) = \frac{1}{t}\left<\phi(\x),\phi(\y)\right>
\end{equation}

Based on IK and kernel mean embedding (KME)~\cite{muandet2017kernel}, IDK is proposed to measure the similarity between two distributions~\cite{ting2023isolation}.
Given the feature mapping $\phi$ of IK and \autoref{eq:IKfm}, the empirical estimation of IDK can be expressed based on the feature mapping $\phi$ of IK $\k$.

\begin{definition}[Isolation Distributional Kernel~\cite{ting2023isolation}]
\label{def:idk}
    For any two distributions $\P_S$ and $\P_T$, where $\P_S$ generates a sample set $S\subset\R^d$, the IDK of $\P_S$ and $\P_T$ is given as:
    \begin{equation*}
    \begin{aligned}
        {\K}(\P_S, \P_T) &= \frac{1}{t|S||T|}\sum_{\x\in S}\sum_{\y\in T} \left<\phi(\x),\phi(\y)\right>\\
        &= \frac{1}{t}\left<{\Phi}(\P_S), {\Phi}(\P_T)\right>,
    \end{aligned}
    \end{equation*}
    where ${\Phi}(\P_S) = \frac{1}{|S|}\sum_{\x\in S}\phi(\x)$ is the kernel mean map of $\P_S$ with $\K$, or the feature map of IDK.
\end{definition}

\subsubsection{Point Anomaly Detector $\K$}

Given a dataset \(D\), we summarize it by a single mapped point \(\widehat\Phi(\mathcal{P}_D)\).
To detect point anomalies, each \(x\in D\) is mapped to \(\widehat\Phi(\delta(x))\), where $\delta(\cdot)$ is a Dirac measure, and its similarity with respect to \(\widehat\Phi(\mathcal{P}_D)\) is computed as
\[
   \mathcal{K}(\delta(x), \mathcal{P}_D) = \langle \widehat\Phi(\delta(x)),\,\widehat\Phi(\mathcal{P}_D)\rangle.
\]
Sorting these similarities yields a ranking of all points in \(D\); anomalies are those points with the smallest similarity to \(\widehat\Phi(\mathcal{P}_D)\).

\subsection{Top-\(k\) Anomalies and Set of Normal Instances}
We sort anomaly scores \(\{a_i\}\) in ascending order and select the \(k\) instances having the smallest scores as the \emph{Top-\(k\)} anomalies.  
The remaining \(n-k\) instances are partitioned into a ``gap" subset consisting of the next 20\% lowest-scoring points and an 80\% ``normal" subset, providing a graded contrast for explanation.

\subsection{Instance-Level Explanation via SiNNe}

\subsubsection{Isolation via Nearest‐Neighbor Ensembles (iNNE) and SiNNe Outlying Score}

SiNNe \cite{samariya2020new} constructs an ensemble of $t$ anomaly indicator models $\mathcal{H}_i$ ($i=1,\dots,t$), each trained on a random subset $\D_i\subset D$ of size $|\D_i|=\psi$.
Each model $\mathcal{H}_i$ defines a \emph{normal region} as the union of all hyperspheres, each centered at $x\in \D_i$, where the radius of the hypersphere centered at $x$ is the Euclidean distance to its nearest neighbor in $\D_i$.
The complement of all of these hyperspheres in $\mathbb{R}^d$ is treated as the \emph{anomaly region}.

A query point $q$ is scored by each model $\mathcal{H}_i$ as
\[
s\bigl(q\mid \mathcal{H}_i\bigr)
=\begin{cases}
1, &\quad\text{if }q\text{ lies outside all hyperspheres of }\mathcal{H}_i\\
0, &\quad\text{otherwise}
\end{cases}
\]
so that $s(q\mid\mathcal{H}_i)=1$ indicates an anomalous decision by model $\mathcal{H}_i$.

The final SiNNe outlying score of $q$ is defined as the scores averaged over the ensemble:
\[
\mathrm{SiNNe}(q|D)
=\frac{1}{t}\sum_{i=1}^{t} s\bigl(q\mid \mathcal{H}_i\bigr),
\]
where a higher $\mathrm{SiNNe}(q)$ denotes a stronger consensus among the ensemble that $q$ is anomalous.

SiNNe  identifies compact subspaces of features that best explain why a query point \(q\) is anomalous or normal. 

Let \(\mathcal{F} = \{f_1, f_2, \dots, f_{d'}\}\) be the complete set of features after preprocessing, where \(d'\) is the number of features. We consider candidate subspaces $S \subseteq \mathcal{F}$, where each \(f \in S\) is a feature that corresponds to a gene.
To ensure interpretability, we perform a search 
for feature subspaces up to depth
$\ell$
for both anomaly and normal explanations, described in the following paragraphs.

\paragraph{Using SiNNe for Anomaly Explanation:}  
Given a detected anomaly \(\textbf{a}\) and a dataset \(D\), the gene subspace $S^*_{\mathrm{anom}}$ found by SiNNe that differentiates $\textbf{a}$ from the cells in $D$ is given as  

\[
S^*_{\mathrm{anom}} = \text{Expl}_{\mathrm{anom}}(\mathbf{a}, D) = \argmax_{\substack{S\subseteq \mathcal{F},\,|S|\le \ell}}\mathrm{SiNNe}(\mathbf{a}_S|D_S)
\]

where $\mathrm{SiNNe}(\textbf{a}_S|D_S)$ denotes the examination of anomaly $\textbf{a}$ with respect to dataset $D$, performed within subspace $S$.

We perform a beam search with a width of \(b=5\) and maximum depth of \(\ell=3\):
the search is initialized with the $b$ individual genes having the highest anomaly scores;
these candidaes are then iteratively expanded by an additional gene, where only the top $b$ subspaces are retained, until the optimal subspace of genes is selected at the final depth.

\paragraph{Using SiNNe for Normal Explanation:}  
Given a set of normal cells $N$ and a typical normal cell $\textbf{z}$ in it, the gene subspace $S^*_{\mathrm{norm}}$ found by SiNNe that characterizes the normal cells is given as:  

\[
S^*_{\mathrm{norm}} = \text{Expl}_{\mathrm{norm}}(\mathbf{z}, N) = \argmin_{\substack{S\subseteq \mathcal{F},\,|S|\le \ell}} \mathrm{SiNNe}(\mathbf{z}_S|N_S)
\]

The same beam search strategy used for anomaly explanation is employed here;
however, in this context, the lowest-scoring subspace is selected as optimal.  

In summary, the improved SiNNe can be applied in two distinct scenarios: first, to explain the anomalousness of an individual cell 
$q$, and second, to characterize a typical cell from a set of normal cells. The original SiNNe \cite{samariya2020new} can only be applied to explain an anomaly.
By limiting \(|S|\le3\) and \(b = 5\) (the default setting), the resulting gene subspaces can then be visualized via 2D/3D plots.

\paragraph{Restricting to Non‐Zero-Expressed Genes for Normal Explanation}  
When explaining a typical normal cell \(\textbf{z}\), we observe that including genes with zero expression in \(\textbf{z}\) trivializes the search:
almost all cells with zero-expressed genes are selected—masking the truly informative (or non-zero expressed) genes.
Therefore, to show what makes \(\textbf{z}\) typical, we first extract a subspace of non-zero-expressed genes:
\[
\F' = \{\,f_j \mid \textbf{z}_{\mathrm j} > 0\}, 
\]  
and restrict our beam search to attribute subspaces \(S\subseteq\F'\).
By focusing only on genes that actually expressed in \(\textbf{z}\), the resulting subspace \(S^*_{\mathrm{norm}}\) highlights those non‐zero markers that anchor \(\textbf{z}\) at the center of its cohort.
In a 2D/3D visualization of the subspace \(S^*_{\mathrm{norm}}\), \(\textbf{z}\) is expected to be located at the center of the cluster core, validating both its typicality and the explanatory power of the selected biologically meaningful genes.

\paragraph{Constructing Reference Set \(\C_\text{Cell-Type}\) of an Anomaly} 

For each anomaly \(\textbf{a}\), we first identify its top-10 nearest neighbors from all the normal cells based on the similarity scores as measured by IDK.
We then determine the majority labels among these 10 neighbors.
Let \(y^*\) denote the class label of the majority.
We define the reference set \(\C_\text{Cell-Type}\) as all samples that share the same label \(y^*\), \textit{i.e.},  
\[
\C_\text{Cell-Type} = \{x \in D\ \mid \text{Cell-Type}(x) = y^*\}.
\] 

This \(\C_\text{Cell-Type}\) is the closest semantically coherent normal class with respect to the anomaly \(\textbf{a}\).
By comparing \(\textbf{a}\) to \(\C_\text{Cell-Type}\), the feature subspace explanation identified is relevant to its closest normal cluster.
The selection of an appropriate reference set is critical for generating diagnostically relevant explanations.
Contrasting an anomaly with a semantically distant normal cluster often identifies high-variance features that, while significant, are diagnostically trivial.
For instance, comparing a cancerous lung cell to its healthy counterpart isolates salient pathological markers. 
Conversely, comparing it to a healthy liver cell would be confounded by fundamental tissue-specific differences, obscuring the actual drivers of malignancy. 
Therefore, contextualizing an anomaly against its immediate normative counterpart is essential to ensure the identified feature subspace is genuinely characteristic of the anomaly.

\subsection{Extending to Group Anomaly Explanation}

The framework presented in this paper is primarily focused on point anomaly detection and explanation.
However, the same principles can be extended to explain group anomalies.
The key difference lies in the anomaly entity (a single cell vs. a group of cells) and the corresponding set for comparison.
\autoref{tab:point_vs_group} outlines the functions used to generate explanations for both point and group anomalies.

\setlength{\aboverulesep}{0pt}
\setlength{\belowrulesep}{0pt}
\begin{table}[h]
\centering
\caption{
framework for point and group anomaly explanations. 
For a point anomaly \(\textbf{a}\), the explanation is derived by comparing it against its closest normal cells \(\C_\text{Cell-Type}\). 
For a group anomaly \(G\), two explanations are provided: (i) to explain how the representative of the group anomaly \(\textbf{r}_G\) differs from representatives of the normal groups \(R_\mathcal{N}\); and (ii) to explain the characteristics of \(\textbf{r}_G\) with respect to its own group \(G\). 
Symbols can be found in \autoref{tab:notation}.
}
\label{tab:point_vs_group}
\renewcommand{\arraystretch}{1.8} 
\begin{tabular}{c|cc|c}
\toprule
& \textbf{Anomaly} & \tabincell{c}{\textbf{Set of closest}\\\textbf{normal cells/groups}}& \textbf{Explanation Function} \\
\midrule
\tabincell{c}{\textbf{Point Anomaly}\\\textbf{Detection}} & 
 \(\textbf{a}\) & \(\C_\text{Cell-Type}\) &$\text{Expl}_{\mathrm{anom}}(\mathbf{a} , \mathcal{C}_{\text{Cell-Type}})$  \\
\midrule
\multirow{3}{*}{\tabincell{c}{\textbf{Group Anomaly}\\\textbf{Detection}}} & 
\(G\) & $\mathcal{N}$ & -
\\  %
& \(\textbf{r}_G\) & \(R_\mathcal{N}\) &
$\text{Expl}_{\mathrm{anom}}(\mathbf{r}_G , R_{\mathcal{N}})$ \\  %
& \(\textbf{r}_G\) & \(G\) &
$\text{Expl}_{\mathrm{norm}}(\mathbf{r}_G , G)$ \\

\bottomrule
\end{tabular}
\end{table}

\subsection{Metric for Comparing Feature Sets and Detectors}

We adopt a metric \(\Gamma(B \!\mid\! A)\) proposed by Ting et al. \cite{Ting2025} that was originally used to compare the relative goodness of two detectors.
Here we assess the relative goodness of two types of feature sets using the same detector,
in this case, the Isolation Distributional Kernel (IDK).

Consider an anomaly detector trained on a dataset of \(n\) instances using either feature set \(A\) or \(B\). 
First, we establish a reference ranked list of anomalies by identifying the top $e$ anomalies using feature set $A$, where $e=\lceil \varepsilon n\rceil=\lceil 0.05 n\rceil\ll n$.
This reference set is denoted as:
\[
\mathcal{T}_A = [q_1, q_2, \ldots, q_e].
\]  
Next, we rank all $n$ instances using feature set \(B\) with the same anomaly detector to produce the ordered list  
$\mathcal{L}_B = [\ell_1, \ell_2, \dots, \ell_n)]$.
We then define the metric $\Gamma(B\!\mid\!A)$ as the minimum rank $r$ required to recover all anomalies from $\mathcal{T}_A$ within the list $\mathcal{L}_B$, normalized by the dataset size:
\[
\Gamma(B\!\mid\!A)
\;=\;
\frac{1}{n}
\min\bigl\{r \;\big|\; \mathcal{T}_A \subseteq [\ell_1,\ell_2,\dots,\ell_r]\bigr\}.
\]
In essence, $\Gamma(B\!\mid\!A)$ is the fraction of the dataset, when ranked by features $B$, that must be examined to find all reference anomalies defined by features $A$.
The symmetric quantity $\Gamma(A\!\mid\!B)$ can be computed analogously by reversing the roles of $A$ and $B$.

Three possible outcomes are:
\begin{itemize}
    \item  \(\Gamma(B\!\mid\!A) \approx \Gamma(A\!\mid\!B) \approx \frac{e}{n} \):
    This outcome indicates a strong agreement between the top-ranking anomalies identified by feature sets $A$ and $B$.
    It is sufficient to use either one of them.
    
    \item  \(\Gamma(B\!\mid\!A) \gg \Gamma(A\!\mid\!B)\):
    This result implies that feature set \(A\) recovers the anomalies prioritized by \(B\), while the reverse is not true, suggesting that feature set $A$ is more comprehensive or powerful, making it a superior choice over $B$.
    
    \item  \(\Gamma(B\!\mid\!A) \approx \Gamma(A\!\mid\!B) \gg \frac{e}{n} \):
    This indicates that feature sets $A$ and $B$ identify distinct and complementary sets of anomalies, as neither can recover the anomalies found by the other.
    This suggests that both $A$ and $B$ should be used in conjunction to achieve the most thorough detection.
\end{itemize}

\section{Authors' contributions}
K.M.T. and D.S. conceived the study and designed the methodological framework. 
D.S. and X.Z. conducted the experiments and performed the data analysis. 
X.L. developed and provided the improved software code used for the analysis. 
D.S., K.M.T., J.Z., and X.L. contributed to the writing of the manuscript. 
K.M.T. and J.Z. supervised the project. 
All authors read and approved the final manuscript.

\section{Funding}
K.M.T. was supported by the National Natural Science Foundation of China (NSFC) under grant numbers W2531050 and 92470116.

\section{Data availability}
All datasets analyzed in this study are publicly available. The Tutorial dataset (1\% Jurkat cells) is available from the scCAD GitHub repository (\url{https://github.com/xuyp-csu/scCAD/blob/main/1%25Jurkat.h5}.
The human Airway dataset is available from the Gene Expression Omnibus (GEO) under accession number GSE103354.
The preprocessed human Tonsil dataset (study SCP2169) and the Crohn's disease dataset (study SCP359) are available from the Broad Institute Single Cell Portal at
(\url{https://singlecell.broadinstitute.org/single_cell/study/SCP2169/slide-tags-snrna-seq-on-human-tonsil}) and (\url{https://singlecell.broadinstitute.org/single_cell/study/SCP359/ica-ileum-lamina-propria-immunocytes-sinai}).

\section{Code availability}
The source code for the explainable rare cell identification framework is publicly available on GitHub at \url{https://github.com/2002sd/Explainable-Rare-Cell-Identification}. 

The implementations of the core algorithms used in this study, Isolation Distributional Kernel (IDK) and improved SiNNe, were based on their original publications. The original source code for these methods can be found at their respective repositories. 

\clearpage

\bibliographystyle{IEEEtran}
\bibliography{reference}

\end{document}